\documentclass{JHEP3}
\usepackage{amsmath}
\usepackage{amsthm}
\usepackage{amsfonts}
\usepackage{amssymb}

\def \Fig#1#2#3 {
\begin{figure}
\centering
\epsfxsize=#2cm \epsfbox{#1.eps}
\caption{#3}
\label{#1}
\end{figure}
}

\newcount\figno
\def\fig#1#2#3{
\par\begingroup\parindent=0pt\leftskip=1cm\rightskip=1cm\parindent=0pt
\baselineskip=15pt
\global\advance\figno by 1
\epsfxsize=#3
\centerline{\epsfbox{#2}}
\vskip 12pt
{\bf \small Figure \the\figno:} {\small #1}\par
\endgroup\par
}
\def\figlabel#1{\xdef#1{\the\figno
\mbox{ }}}
\def\encadremath#1{\vbox{\hrule\hbox{\vrule\kern8pt\vbox{\kern8pt
\hbox{$\displaystyle #1$}\kern8pt}
\kern8pt\vrule}\hrule}}

\newcommand{\OSP}{\text{OSP(1$|$2)}}
\newcommand{\osp}{\text{osp(1$|$2)}}
\newcommand{\tsl}{\text{sl$_2$}}
\newcommand{\tOSP}{\text{OSP}}
\newcommand{\tWZNW}{\text{WZNW}}
\def\Hrep{{}}
\title{Structure constants of the \OSP\ WZNW model}
\author{Yasuaki Hikida and Volker Schomerus\\
DESY Theory Group, DESY Hamburg, Notkestrasse 85, D-22603 Hamburg, Germany\\
E-mail: \email{yasuaki.hikida@desy.de}, \email{volker.schomerus@desy.de}
}


\abstract{We propose exact formulas for the 2- and 3-point functions of
the WZNW model on the non-compact supergroup \OSP. Using the path
integral approach that was recently developed in arXiv:0706.1030
we show how local correlation functions in the \tOSP($p|$2) WZNW
models can be obtained from those of ${\cal N}=p$ supersymmetric
Liouville field theory for $p=1,2$. We then employ known results
on correlators in ${\cal N}=1$ Liouville theory to determine the
structure constants of the \OSP\ theory.
}

\preprint{ arXiv:0711.0338 \\ DESY 07-190 }

\begin{document}





\section{Introduction}

Two dimensional conformal field theories with target space
supersymmetry have recently been studied intensively because of
the important role they play for various problems ranging from
disordered electron systems to string theory. Through the AdS/CFT
correspondence, for instance, 4-dimensional theories with
superconformal symmetry are related to closed strings moving in a
target space with psl(2,2$|$4) symmetry (see e.g.\ \cite{Metsaev:1998it}
for a concrete world-sheet model). In condensed matter theory,
on the other hand, methods have been developed \cite{Efetov} which
permit the computation of spectral densities, transport properties,
etc., in systems with random disorder. These involve auxiliary
field theories with internal supersymmetry, in particular models
with central charge $c=0$ whose action is invariant under global
osp(2N$|$2N) transformations (see e.g.\ \cite{Bernard} for a
review and further references).
\smallskip

Conformal field theories on superspaces provide a rich class of
non-unitary and non-rational models. Moreover, their correlators
often possess logarithmic singularities, a property that was
explained through harmonic analysis on supergroups in
\cite{SchSal}. Given all these features, models with target space
supersymmetry may appear rather difficult to solve! Some progress
was made recently for a large class of basic Lie superalgebras. In
fact, the solution of WZNW models on type I supergroups has been
reduced to solving an associated bosonic WZNW model. The type I
condition (see below for more details) is satisfied for many
interesting examples, in particular for the supergroups
PSL(N$|$N), but it excludes almost all supergroups \tOSP(M$|$2N),
with the exception of \tOSP(2$|$2N). Therefore, the solution of
WZNW models on \tOSP(M$|$2N) remains an important open problem. In
this note we shall develop a new approach to the issue and we
shall employ it to solve the WZNW model on \OSP. \medskip

For a moment we shall put the topic of sigma models on superspaces
aside and talk about a somewhat unrelated development in the area
of non-rational conformal field theory (CFT). The two most
thoroughly studied examples of CFTs with non-compact target space
are Liouville theory and the WZNW model on the 3-dimensional
hyperboloid $H^+_3$, a Euclidean version of $AdS_3$. The structure
constants of Liouville field theory were first proposed by Dorn,
Otto and the Zamolodchikovs
\cite{Dorn:1994xn,Zamolodchikov:1995aa}. The original proposal has
meanwhile been established rigorously through a series of papers
by Teschner, partly in collaboration with Ponsot
\cite{Ponsot:1999uf,Teschner:2001rv,Teschner:2003en}. At the same
time, Teschner also extended the successful solution of Liouville
theory to the $H^+_3$ model
\cite{Teschner:1997ft,Teschner:1999ug}. It turned out that the
structure constants of Liouville theory appear as building blocks
for those of the WZNW model. This may not come entirely unexpected
since it is often stated that Liouville theory provides a model
for the radial direction of $AdS$. In any case, the close relation
between the two theories has furnished rather useful tools, e.g.\
to prove consistency of structure functions in the $H^+_3$ model
\cite{Teschner:2001gi}. A precise relation between local
correlation functions of the $H^+_3$ WZNW model and Liouville
field theory on the complex plane was later put together by
Ribault and Teschner \cite{Ribault:2005wp}, based on earlier 
related works \cite{Sklyanin:1987ih,Stoyanovsky}.
\smallskip

In our recent paper \cite{HS} we revisited and extended the
relation between the WZNW model on $H^+_3$ and Liouville field
theory. Most importantly, we rederived the relation found in
\cite{Ribault:2005wp} very naturally using a path integral
formalism. As one application of our new approach we then
generalized the correspondence to correlation functions on
arbitrary closed surfaces. The path integral treatment 
clearly suggests that similar correspondences may exist
for other models with an \tsl\ current algebra symmetry. 
Furthermore, whenever this is the case, the path integral 
approach provides concrete tools to determine the precise 
relation between local correlation functions of two models. 
We shall often refer to such a correspondence as a
`reduction' simply because it expresses correlation functions of a
model on a $d$-dimensional target space through correlators in
another local field theory with target space of dimension $d-2$.
Let us stress, however, that the correspondence goes far beyond
the well known Hamiltonian reduction (see \cite{Bouwknegt:1992wg}
for a review and references to early contributions). Most importantly, 
the new correspondence imposes absolutely no restrictions on the momenta
of the tachyon vertex operators, in spite of the difference
between target space dimensions. While target space momenta in the
original theory with \tsl\ symmetry possess $d$ parameters, only
$d-2$ components are needed to parametrize momenta in the reduced
model. The correspondence stores all information about the missing
two momentum components through a highly intriguing mechanism in
the {\em world-sheet} location of additional degenerate field
insertions.
\medskip

Let us now connect the correspondence between the $H^+_3$ WZNW
model and Liouville theory (or its generalizations) to the main
goal of this work, namely the solution of WZNW models on \tOSP\
supergroups. All classical Lie superalgebras possess a so-called
distinguished $\mathbb{Z}$-gradation in which fermionic generators
possess degree $\delta = \pm 1$. For type I superalgebras, all
bosonic generators are located in degree zero. Because there exist
no bosonic elements with degree $\delta = +2$, anti-commutators
between fermionic elements of degree $\delta = +1$ vanish. This
feature of type I superalgebras was exploited in \cite{QuSc} and
leads to a solution of the corresponding WZNW models. For type II
superalgebras the situation is different. By definition, these
contain bosonic generators of degree $\delta = \pm 2$ with respect
to the distinguished $\mathbb{Z}$-gradation. Consequently,
anti-commutators between fermionic elements of degree $\delta =
+1$ need no longer vanish, a property that spoils a successful
solution along the lines of \cite{QuSc}. Our idea here is to
remove the problematic generators of degree $\delta = \pm 2$
through reduction to a local field theory with a lower dimensional
target space. Note that each generator of degree $\delta = +2$
determines a unique \tsl\ subalgebra within the current
superalgebra of a type II WZNW model. Moreover, the different
\tsl\ current algebras that are obtained in this way commute with
each other. Therefore, we can apply the `reduction' outlined in
the previous paragraph to each of the \tsl\ algebras, one after
another. In this note we shall restrict ourselves to examples with 
a single bosonic generator of degree $\delta = +2$. In fact, most 
of our analysis focuses even further to the example of \osp.
\medskip

Let us now outline the main results of this paper and describe the
content of each of the following sections. We shall begin in
section 2 by spelling out the action of the \OSP\ WZNW model in a
first order formulation. The Lie superalgebra \osp\ contains a
bosonic \tsl\ subalgebra along with two fermionic generators.
After `reduction' there remains a single bosonic direction and two
fermionic ones. The latter are shown to provide the fermionic
partner of the bosonic field in ${\cal N}=1$ Liouville field theory and
an additional free fermion. Thereby, we shall relate correlators
in the \OSP\ WZNW model to those in a product of ${\cal N}=1$ Liouville
theory with a free fermion model. We shall then briefly
demonstrate how our strategy extends to higher dimensional OSP
supergroups. In particular, we relate the \tOSP(2$|$2) WZNW model
to ${\cal N}=2$ Liouville theory.\footnote{As we remarked before, the
supergroup \tOSP(2$|$2) $\cong$ SL(1$|$2) is of type I and
therefore it can be dealt with along the lines of \cite{QuSc}. Our
treatment here shall be based on a natural $\mathbb{Z}$-gradation
in which \tOSP(2$|$2) possesses a single bosonic generator in
degrees $\delta =  \pm 2$ (i.e.\ not the distinguished
$\mathbb{Z}$-gradation). This makes it an example for our new
approach and thereby provides an alternative way of solving the
model, different from the one outlined in \cite{QuSc}.} The
relation between osp$(p|2)$ current algebras and ${\cal N}=p$
superconformal symmetries is not new. In fact, it was known for a
long time how to obtain the latter from the former through
Hamiltonian reduction \cite{Bershadsky:1989tc}. But let us stress
once more that our correspondence goes much beyond a mere
reduction since it establishes an equivalence between correlators
of the two theories.
\smallskip

The rest of the paper is then devoted to the computation of 2- and
3-point functions in the \OSP\ WZNW model. In section 3 we shall
study the minisuperspace approximation to the structure constants.
This will also allow us to gain some experience with the \osp\
invariant tensors which appear as building blocks for the particle
and field theory quantities alike. Section 4 contains formulas for
the structure constants of the \OSP\ WZNW model. The 2-point functions
of the theory can be determined easily from the known 2-point
functions of ${\cal N}=1$ Liouville field theory. The 3-point functions
of the WZNW model are related to certain 4-point functions in the
reduced model. Luckily, the relevant 4-point correlators in ${\cal N}=1$
Liouville theory have been constructed in \cite{FH}. When
combined with the appropriate correlation functions in free
fermionic field theory, the resulting 3-point function of the WZNW
model becomes manifestly \osp\ invariant and the structure
constants can be read off.
\smallskip

Our final formula for the 3-point correlator will involve tachyon
vertex operators $V^j(x,\xi|z) = V^j(x,\bar x,\xi, \bar \xi|z)$
which are labeled by a spin $j$, a complex coordinate $x$ and 
a complex Grassmann variable $\xi$. From time to time we shall 
also display the dependence on the complex conjugate variables
$\bar x$ and $\bar \xi$ in order to show that the corresponding 
quantities are not chiral. The parameters $j,x,\xi$ are chosen 
such that the operator products with supercurrents $J_X$ of the 
\OSP\ WZNW model take the form
$$ J_X(z) \ V^{j} (x,\xi|w) \ \sim \ r_X V^{j}
(x,\xi|w)\frac{1}{z-w} + \dots \ \ . $$ Here, the subscript $X$ runs
through a basis $X = E^\pm, F^\pm, H$ in \osp\ (see appendix
\ref{App:osp} for details on \osp). The symbols $r_X$ on the right
hand side of the operator product denote certain first order
differential operators acting on $x, \bar x$ and $\xi,\bar \xi$,
see eqs.\ \eqref{rlist}. The 3-point function of tachyon vertex
operators reads
\begin{align}\nonumber
 &\langle V^{j_1} (x_1,\xi_1|z_1) V^{j_2} (x_2,\xi_2|z_2)
          V^{j_3} (x_3,\xi_3|z_3) \rangle  \\[2mm]
 & \hspace*{3cm} = \  \frac{C^\Hrep_b (j_1,j_2,j_3) +  \tilde
     C^\Hrep_b (j_1,j_2,j_3) \eta \bar
   \eta}
   {| X_{12} |^{-2 j_{12}-1}
            | X_{23} |^{-2 j_{23} -1} | X_{31} |^{-2
            j_{31}-1}} \
            \frac{1}{\prod_{i < j}|z_{ij}|^{2 \Delta^\Hrep_{ij}}}
\label{3ptx}
\end{align} where $z_{ij} = z_i-z_j$, $j_{12} = j_1 + j_2 - j_3$ etc.\  
and $X_{ij}= x_i-x_j-\xi_i\xi_j$. The exponents $\Delta_{ij}$ are 
determined by the conformal dimensions $$\Delta^\Hrep_j = 
-2b^2 (j+1)(j+\tfrac12) \ \ \ \ \mbox{where} \ \ b^{-2} \ = \ 2k-3$$ 
through $\Delta_{12} = \Delta_{j_1} + \Delta_{j_2} - \Delta_{j_3}$ etc. 
An explicit formula for the super-projective 3-point invariants $\eta,\bar 
\eta$ is given in eq.\ \eqref{eta}. The form of the 3-point functions is 
determined by world-sheet conformal
symmetry and target space \osp\ invariance up to the two functions
$C^\Hrep_b$ and $\tilde C^\Hrep_b$. Expressions for these are provided
in eqs.\ \eqref{tC} and \eqref{C} at the very end of this note.
Thereby, the non-rational \OSP\ WZNW model is solved. Structure 
constants for a compact target space may be obtained by analytic 
continuation of the momenta. Such models have been argued to 
describe the continuum limit of certain super-spin chains, 
see \cite{Saleur:2003zm} and \cite{Essler:2005ag} for the 
cases of \osp\ and osp(2$|$2), respectively. An OSP(2$|$2) 
WZNW model also emerges in the study of $2+1$ dimensional 
spin-full electrons with random gauge potential, see 
\cite{Bernard:2000vc,Bhaseen:2000mi,LeClair:2007aj} and 
further references therein.

\section{Supergroup models and super Liouville field theory}
\label{secSRT}

The aim of this section is to derive a relation between the
OSP($p|$2), $p=1,2,$ WZNW model and the product of a supersymmetric
Liouville theory with a theory of $p$ free fermions. Let us note
that the supergroup OSP($p|$2) has superdimension {\it sdim} OSP($p|$2)  
$=(\tfrac12 (p^2-p)+3)|2p$. The manipulations to be carried out in the 
current section work for all $p$. They relate the WZNW model to a new
interacting field theory on a target space of superdimension
$(\tfrac12 (p^2-p)p+1)|p$ and an additional model 
of $p$ free massless fermions. Two bosonic directions are integrated 
out explicitly while half of the fermions turn out to decouple. When 
$p=1,2$, the field content of the interacting sector is that of 
${\cal N}=1,2$ Liouville field theory and we shall see that the 
actions also agree. For larger values of $p$, the corresponding 
lower dimensional model has not been studied before so that the 
relation is of limited use. For this reason, we shall mostly 
focus on the case of $p=1$ and then spell out the relation for 
$p=2$. Larger values of $p$ may be treated in the same way.

\subsection{\OSP\ WZNW model from ${\cal N}=1$ Liouville theory}

In this subsection, we focus on the simplest example and derive
the relation between correlators of \OSP\ WZNW model and
${\cal N}=1$ super Liouville field theory. After a few
introductory comments on the action of the \OSP\ WZNW model,
we shall pass to a first order formulation involving two
additional bosonic auxiliary fields along with two fermionic ones.
Following the ideas of \cite{HS}, we can then integrate out four
bosonic fields. The resulting theory contains a single bosonic
field and two pairs of chiral fermions. Their action is finally
rewritten as a sum of an ${\cal N}=1$ Liouville model and a free fermion
theory.%
\smallskip

For any (super-)group, the action of WZNW model takes the
following standard form,
\begin{align}
 S^{\tWZNW} (g) \ = \ \frac{k}{4 \pi} \int_{\Sigma} d ^2z
 \langle g^{-1} \partial g , g^{-1} \bar \partial g \rangle
 + \frac{k}{2 4 \pi} \int_{B} \langle g^{-1} d g ,
  [g^{-1} d g , g^{-1} d g ] \rangle
  \label{WZNWaction}
\end{align}
where the integrations are over a world-sheet $\Sigma$ and a three
dimensional manifold with $\partial B = \Sigma$, respectively. The
Lie superalgebra \osp\ has superdimension $3|2$ with bosonic and
fermionic  generators denoted by $E^\pm, H$ and by $F^\pm$,
respectively. Their (anti-)commu\-tation relations may be found in
appendix \ref{App:osp} along with explicit formulas for the
metric we use.
\smallskip

We shall adopt a specific parametrization of elements $g \in$ \OSP\
by splitting them into a product $g = \alpha G \beta$ of three
$3\times 3$ supermatrices which are defined by
\begin{align}
  \alpha &=\ e^{2 \theta F^+} ~,
    &\beta &=\  e^{2 \bar \theta F^-}
 &G &= \
 \begin{pmatrix}
  g_B & 0 \\
  0 & 1
 \end{pmatrix} ~,
 &g_B &=\
 \begin{pmatrix}
  1 & \gamma \\
  0 & 1
 \end{pmatrix}
 \begin{pmatrix}
  e^{\phi} & 0 \\
  0 & e^{- \phi}
 \end{pmatrix}
 \begin{pmatrix}
  1 & 0 \\
  \bar \gamma & 1
 \end{pmatrix} ~.
 \label{para1}
\end{align}
The action of the WZNW model can now be spelled out explicitly in
terms of three bosonic fields $\phi,\gamma,\bar \gamma$ and two
fermionic ones $\theta,\bar\theta$. To this end, we decompose the
elements $g = \alpha G \beta$ into its three factors and then
split the WZNW action \eqref{WZNWaction} into several terms with
the help of the Polyakov-Wiegmann identity,
\begin{align}
  S^{\tWZNW} (\alpha G \beta)  &= \ S^{WZNW} (G)
 + \frac{k}{2 \pi} \int d^2 z
   \langle \alpha^{-1} \bar \partial \alpha , \partial G G^{-1} \rangle
 \nonumber \\[2mm]
 & \ \ \ + \ \frac{k}{2 \pi} \int d^2 z
   \langle G^{-1} \bar \partial G , \partial \beta \beta^{-1} \rangle
 + \frac{k}{2 \pi} \int d^2 z \langle \alpha^{-1} \bar \partial \alpha
   , G \partial \beta \beta^{-1} G^{-1} \rangle ~. \nonumber
\end{align}
Inserting our parametrization of the factors $\alpha,\beta$ and
$G$, we obtain the following formulas for the action of the
\OSP\ WZNW model
\begin{align}
  S^{\tWZNW} (g)
   = \frac{k}{2\pi} \int d^2 z \left[ \partial \phi \bar \partial \phi
     + e^{-2 \phi}
    (\partial \bar \gamma -  \bar \theta \partial \bar \theta )
    ( \bar \partial \gamma - \theta \bar \partial \theta )
     + 2 e^{-\phi} \bar \partial \theta \partial \bar \theta  \right] ~.
     \label{osp12}
\end{align}
Note that the theory is interacting with terms up to forth order
in the fermionic fields. For WZNW models on type I supergroups a
special parametrization could be found \cite{QuSc} in which the
interaction terms are at most quadratic in the fermionic fields.
It is a basic feature of the type II case that such a
simplification cannot be achieved.
\smallskip

In order to apply the method of \cite{HS}, it is essential to
change the action in the first order formulation. Introducing new
bosonic variables $\beta, \bar \beta$ as well as the fermionic
ones $p ,\bar p$, all four of weight $\Delta = 1$, the action
may be rewritten as
\begin{align}
  S^{\tWZNW} (g)
  \ = \ \frac{1}{2\pi} \int d^2 z \, \Bigl[ \, \frac12 \partial \phi \bar \partial \phi
   &+ \frac{b}{8} \sqrt g R \phi
 + \beta ( \bar \partial \gamma - \theta \bar \partial \theta)
 + \bar \beta ( \partial \bar \gamma - \bar \theta \partial \bar \theta)
 \nonumber \\[2mm]
 &+ p \bar \partial \theta + \bar p \partial \bar \theta
 - \frac{1}{k} \beta \bar \beta e^{2b \phi}
 - \frac{1}{2k} p \bar p e^{ b \phi}\, \Bigr]
 \label{WZNW}
\end{align}
with $b = 1/\sqrt{2 k - 3}$. Before we continue studying this
theory let us briefly convince ourselves that the original model
agrees with the first order formulation we propose. The equations
of motion for new auxiliary fields read
\begin{align}
 \beta &=\ k e^{- 2b \phi} ( \partial \bar \gamma - \bar \theta \partial \bar \theta )~,
  &\bar \beta &=\ k e^{- 2b \phi} ( \bar \partial  \gamma -
  \theta \bar \partial \theta ) ~,\\[2mm]
  p &=\ - 2 k e^{- b \phi} \partial \bar \theta ~,
 &\bar p &= \ 2 k e^{- b \phi} \bar \partial \theta ~.
\end{align}
Inserting these expressions into the action \eqref{WZNW} we
reproduce the original action \eqref{osp12} apart from the
additional linear dilaton term that appears in eq.\ \eqref{WZNW}.
The latter arises from the Jacobian in the change of variables.
Utilizing the formulas
\begin{align}
  \ln \det (A \partial B \bar \partial)
 = \frac{1}{48 \pi} \int d^2 z
  ( | \partial \ln A |^2 + | \partial \ln B |^2
   - 4 \partial \ln A \bar \partial \ln B )
\end{align}
and $\sqrt g R = - 4 \partial \bar \partial \ln |\rho|^2$
for the world-sheet metric $ds^2 = |\rho|^2 dz d \bar z$, we
obtain the following bosonic contribution
\begin{align}
 - \ln \det (| \rho |^{-2} e^{2 \phi} \partial e^{-2 \phi} \bar \partial)
= - \frac{1}{\pi} \int d^2 \partial \phi \bar \partial \phi
  + \frac{1}{8 \pi} \int d^2 \phi \sqrt g R \phi~,
\end{align}
along with a fermionic contribution of the same form
\begin{align}
 \ln \det (| \rho |^{-2} e^{\phi} \partial e^{- \phi} \bar \partial)
 = \frac{1}{4 \pi} \int d^2 \partial \phi \bar \partial \phi
  - \frac{1}{16 \pi} \int d^2 \phi \sqrt g R \phi~.
\end{align}
This concludes our derivation of the action \eqref{WZNW} from the
generic formulation \eqref{WZNWaction} of the WZNW model on the
supergroup \OSP.
\medskip

We are now prepared to begin analyzing correlation functions in
the \OSP\ WZNW model. The $N$-point functions of tachyon vertex
operators are given by
\begin{align}
 \langle \prod _{\nu=1}^N V^{\epsilon_\nu,\bar \epsilon_\nu}
  _{j_\nu} (\mu_\nu | z_\nu) \rangle
  &=\  \int  \, {\cal D} \phi {\cal D}^2 \beta {\cal D}^2 \gamma {\cal D}^2 \theta
  {\cal D}^2 p  \  e^{-S ^{\tWZNW}(g)} \ \prod_{\nu=1}^{N} V^{\epsilon_\nu,\bar \epsilon_\nu}
  _{j_\nu} (\mu_\nu | z_\nu)\ .
\end{align}
The vertex operators $V^{\epsilon,\bar \epsilon}_{j}(\mu|z)$ we
inserted in the points $z=z_\nu$ depend on the SL(2,$\mathbb{C}$)
quantum numbers $j,\mu,\bar \mu$ and an additional choice of
$\epsilon,\bar \epsilon = \pm 1$. They are defined by
\begin{align}
V^{\epsilon,\bar \epsilon}_{j}
  (\mu |  z) &=\
 e^{\epsilon \sqrt{\mu} \theta}
  e^{\bar \epsilon \sqrt{\bar \mu} \bar \theta}
 | \mu |^{2j + 2}
 e^{\mu \gamma - \bar \mu \bar \gamma}
  e^{2 b (j +1) \phi} ~ .
\end{align}
The bosonic factor is the same as in \cite{HS}. As one may see by
expanding exponentials, the fermionic factors are sufficient to
generate $1,\theta,\bar \theta$ and $\theta\bar \theta$. The basis
we have chosen, including the factors $\sqrt{\mu}$ and $\sqrt{\bar
\mu}$ in front of the fermionic fields, turn out to be very
convenient for what we are about to do.
\smallskip

Having set up our problem, we proceed along the lines of \cite{HS}
and integrate out $\gamma, \bar \gamma$ first and then $\beta,\bar
\beta$ using the following change of variables
\begin{align} \label{B}
 \sum_{\nu = 1}^N \frac{\mu_\nu}{w - z_\nu}
  = u \frac{\prod_{i=1}^{N-2}(w-y_i)}{\prod_{\nu=1}^N (w-z_\nu)}
  =: u {\cal B} (y_i,z_\nu;w)
\end{align}
and a similar equation for the conjugate variables. This relation
defines the parameter $u$ and the world-sheet coordinates $y_i$ in
terms of $\mu_\nu$. After an appropriate redefinition of the
scalar field $\phi$ (see \cite{HS} for many more details) we
obtain
\begin{align}
  \langle \prod _{\nu=1}^N V^{\epsilon_\nu,\bar \epsilon_\nu}_{j_\nu}
  (\mu_\nu | z_\nu)  \rangle
  &=\  \delta^2 (\sum_{\nu = 1}^N \mu_\nu)\ \, |u|\, |\tilde  \Theta_N|^2
 \, \int \ {\cal D} \varphi  {\cal D}^2 \theta {\cal D}^2 p\
   e^{-S [\varphi,\theta,p]} \ \times \nonumber \\[2mm]
    &\times
   \prod_{\nu=1}^{N}  e^{\epsilon_\nu \sqrt{\mu_\nu} \theta +
    \bar \epsilon_\nu \sqrt{\bar \mu_\nu} \bar \theta}
   e^{2 ( b(j_\nu+1) + \frac{1}{4b}) \varphi} (z_\nu)
    \prod_{j=1}^{N-2} e^{ - \frac{1}{2b} \varphi} (y_j) ~,
\end{align}
where the new action is now given by
\begin{align}
   S [\varphi,\theta,p]
   \ = \ \frac{1}{2\pi} \int d^2 z \, \Bigl[ \, \frac12 \partial \varphi \bar \partial \varphi
   &+ \frac{Q_\varphi}{8} \sqrt g R \varphi
 - u {\cal B} \theta \bar \partial \theta
  + \bar u \bar {\cal B} \bar \theta \partial \bar \theta
 \nonumber \\[2mm]  &
 + p \bar \partial \theta + \bar p \partial \bar \theta
 + \frac{1}{k} e^{ 2 b \varphi}
 + \frac{i}{2k} \frac{1}{|u{\cal B}|} p \bar p e^{ b \varphi}\, \Bigr]\ .
\end{align}
The background charge for the new scalar field $\varphi$ is
shifted from $Q_\phi = b$ to $Q_\varphi = b + 1/b$ and we also
introduced the shorthand
\begin{align}
 \tilde \Theta_N \ = \ \prod_{\mu < \nu}^N
 (z_\mu - z_\nu)^{\frac{1}{4b^2}}
 \prod_{i < j}^{N-2} (y_i - y_j)^{\frac{1}{4b^2}}
 \prod_{\nu = 1}^N \prod_{i=1}^{N-2} (z_\nu - y_i)^{-\frac{1}{4b^2}}  ~.
 \label{ttheta}
\end{align}
As it stands, the action still includes an explicit dependence on
the world-sheet coordinates $z_\nu$ and $y_i$ through the function
${\cal B}$ that we introduced in eq.\ \eqref{B}.
\smallskip

Our next step is to absorb the unwanted factors ${\cal B}$ through
a redefinition of the fermionic fields. In an intermediate step,
we introduce
\begin{align}
 p ' &:= \ p / \sqrt{u {\cal B}} ~,
&\bar p ' &:=\  \bar p /\sqrt{-\bar u \bar {\cal B}} ~,
 &\theta ' &:=\  \theta \sqrt{u {\cal B}} ~,
 &\bar \theta ' &:=\  \bar \theta \sqrt{-\bar u \bar {\cal B}}  ~.
\end{align}
When rewritten through the new fermionic variables, the kinetic
terms become 
\begin{equation}
 - {u {\cal B}} \theta \bar \partial \theta + p \bar \partial \theta
  \ =\  -  \theta ' \bar \partial \theta ' +
    p ' \bar \partial \theta ' -
 \left(p' \theta ' + \frac12 \partial \ln \sqrt{u {\cal B}} \right)
 \bar \partial \ln \sqrt{u {\cal B}}\ \ . \nonumber
\end{equation}
Here, the non-trivial shift from $p'\theta'$ to $p'\theta' + 
\frac12 \partial \ln \sqrt{u{\cal B}}$ is a result of regularization. 
At the same time, the fermionic terms $\exp (\epsilon 
\sqrt{\mu} \theta(z_\nu)) $ get replaced by $\exp (\epsilon 
\theta'(z_\nu))$.%
\footnote{We use a regularization scheme 
$\lim_{w \to z} ( w - z ) = - \ln \rho(z) $ for the
world-sheet metric $ds^2 = |\rho(z)|^2 dz d\bar z$.
For more details, see \cite{HS}.}
Note that the function
$\sqrt{ u \cal B}$ has weight $\Delta = 1/2$ so that after the
redefinition, our new fermionic fields $\theta',\bar \theta', p',
\bar p'$ all possess the same weight $\Delta = 1/2$. We can make
the kinetic terms look more symmetric if we adopt the following
new basis for fermions,
\begin{align}
 \chi &:=  \frac{i}{\sqrt2} ( 2 \theta ' -    p ') ~,
 &\psi &:=  \frac{1}{\sqrt2} p' ~,
 &\bar \chi &:=  - \frac{i}{\sqrt2} ( 2 \bar \theta ' -   \bar p ') ~,
 &\bar \psi &:=  \frac{1}{\sqrt2}  \bar p'  ~.
\end{align}
After inserting these expressions, the chiral kinetic terms read
\begin{equation} \label{kin2}
 - {u {\cal B}} \theta \bar \partial \theta + p \bar \partial \theta
 \ = \ \frac12 \chi \bar \partial \chi
  + \frac12 \psi \bar \partial \psi
  +  \left(i \psi \chi - \frac12\partial \ln \sqrt{u {\cal B}} \right)\bar
\partial \ln \sqrt{u {\cal B}} ~.
\end{equation}
The fermionic contribution $\exp(\epsilon \theta'(z_\nu)) = 1 +
\epsilon \theta'(z_\nu)$ gets replaced by $1+ \epsilon (\psi
-i\chi)/\sqrt{2}$. Note that both the vertex operators and the
term involving $\psi \chi$ in the action mix the two fermions. In
addition, the action still contains terms with $z_\nu$-dependent
coefficients. In order to proceed, we now bosonize the two
fermionic fields $\psi$ and $\chi$,
\begin{align}\label{bos}
 \psi \pm i \chi =\ \sqrt{2} \exp ( \pm i Y) ~, \qquad
 \bar \psi \pm i \bar \chi =\ \pm i \sqrt{2} \exp ( \pm i \bar Y) ~.
\end{align}
The main advantage of this bosonization is that we can now express
the product $\psi \chi$ as a derivative. Thereby, we may now
rewrite the $z_\nu$-dependent terms in the action as follows,
\begin{align}
   & i \psi(w) \chi(w) \bar\partial \ln \sqrt{u {\cal B}}
  \ = \  - i \partial Y(w) \bar \partial \ln \sqrt{u \cal B}  \\[2mm]
   & \quad \sim \ i Y(w) \partial \bar \partial \ln \sqrt{u \cal B}
  \ = \ - \pi i Y(w) [ \sum_{\nu=1}^{N} \delta^2 (w - z_\nu) -
   \sum_{i=1}^{N-2} \delta^2 (w - y_i) ] \ .
\nonumber
\end{align}
The symbol $\sim$ means equality up to total derivatives. In the
new form, we recognize that the corresponding terms only lead to
contributions which are localized at the points $z_\nu$ and $y_i$
of the world-sheet, i.e.\ they modify the vertex operators. Our
final result for the correlation function of tachyon vertex
operators in the \OSP\ WZNW model now reads
\begin{align} 
 \label{H2L}
  &\langle \prod _{\nu=1}^N V^{\epsilon_\nu,\bar \epsilon_\nu}_{j_\nu}
  (\mu_\nu | z_\nu) \rangle
  \ =\  \delta^2 (\sum_{\nu = 1}^N \mu_\nu) | \Theta^{(1)}_N|^2
 \int {\cal D} \varphi  {\cal D}^2 \psi {\cal D}^2 \chi
   \ e^{-S [\varphi,\psi,\chi]}\,  \times\\[2mm]
    & \qquad \times
   \prod_{\nu=1}^{N}
   (e^{\frac{i}{2} Y} + \epsilon e^{- \frac{i}{2} Y} )
   (e^{ - \frac{i}{2} \bar Y} - \bar \epsilon e^{\frac{i}{2} \bar Y})
e^{2 ( b(j_\nu+1) + \frac{1}{4b}) \varphi} (z_\nu)
    \prod_{j=1}^{N-2} e^{- \frac{i}{2} (Y - \bar Y)} e^{
 - \frac{1}{2b} \varphi} (y_j) ~.  \nonumber
\end{align}
The factor $\Theta_N^{(1)}$ combines the function $\tilde \Theta_N$ we had
defined previously in eq.\ \eqref{ttheta} with the numerical
contributions from the second term in eq.\ \eqref{kin2},
\begin{align}
 \Theta^{(1)}_N \ = \ u
 \prod_{\mu < \nu}^N (z_\mu - z_\nu)^{\frac{1}{4b^2}-\frac14}
 \prod_{i < j}^{N-2} (y_i - y_j)^{\frac{1}{4b^2}-\frac14}
 \prod_{\nu = 1}^N \prod_{i=1}^{N-2} (z_\nu - y_i)^{-\frac{1}{4b^2}+\frac14}  ~.
 \label{theta1}
\end{align}
After all our manipulations, the functional $S =
S(\varphi,\psi,\chi)$ is a sum of the action for ${\cal N}=1$
super Liouville theory $(\varphi,\psi,\bar \psi)$ and the action
for a non-interacting real massless fermion $(\chi,\bar \chi)$,
i.e.\
\begin{align}
   S [\varphi,\psi,\chi]
  & = \  \frac{1}{4\pi} \int d^2 z \, \Bigl[\,  \partial \varphi \bar \partial \varphi
   + \frac{Q_\varphi}{4} \sqrt g R \varphi + \frac{2}{k} e^{ 2 b \varphi}\, +
   \nonumber \\[2mm]
 & \hspace*{2cm} +
   \psi \bar \partial \psi + \bar \psi \partial \bar \psi
 + \chi \bar \partial \chi + \bar \chi \partial \bar \chi
 - \frac{2}{k} \psi \bar \psi e^{ b \varphi}\,  \Bigr] ~.
\end{align}
The vertex operators we insert at $z_\nu$ and $y_i$ have been
written in terms of chiral components $Y$ and $\bar Y$ of a
bosonic field. Through the relations \eqref{bos}, we may think of
$Y$ and $\bar Y$ as local functional on the space of fermionic
fields $\psi,\chi$ and $\bar \psi,\bar \chi$. Later, we shall
re-express the relevant exponentials through spin-fields in the
fermionic sector.

\subsection{OSP(2$|$2) WZNW model from ${\cal N}=2$ super Liouville theory}

In the previous subsection, we have shown that the correlators of
\OSP\ model can be expressed in terms of ${\cal N}=1$ super
Liouville field theory with additional fermions. As we have
remarked in the introduction, the ideas work much more generally.
As an example of such generalizations, we shall briefly analyze
the OSP(2$|$2) model. Even though OSP(2$|$2) $\cong$ SL(1$|$2) is
of type I, we shall treat it in the same way as in the \OSP\ model
in the previous section. This amounts to choosing an unusual
$\mathbb{Z}$-grading in which the four fermions are assigned
grades $\delta = \pm 1$ such that there exists one bosonic
generator of grade $\delta = \pm 2$ each. For the readers'
convenience we have listed the anti-commutation relations of the
Lie superalgebra osp(2$|$2) in appendix \ref{App:sl12}.
\smallskip

In order to spell out the action of the WZNW model we need to
adopt specific coordinates on supergroup OSP(2$|$2). Here we
shall use the parametrization $g = \alpha G \beta$ with
\begin{align}
  \alpha &=\ e^{\theta_1 F^+ + \theta_2 \bar F^+} ~,
    &\beta &=\  e^{\bar \theta_1 F^- + \bar \theta_2 \bar F^-} ~ ,
 &G &=\
 \begin{pmatrix}
  g_B & 0 \\
  0 & e^{2i\varphi_1}
 \end{pmatrix} ~,
 \label{paraosp}
\end{align}
where the bosonic part is
\begin{align}
 g_B &=\  e^{i \varphi_1}
 \begin{pmatrix}
  1 & \gamma \\
  0 & 1
 \end{pmatrix}
 \begin{pmatrix}
  e^{\phi_2} & 0 \\
  0 & e^{- \phi_2}
 \end{pmatrix}
 \begin{pmatrix}
  1 & 0 \\
  \bar \gamma & 1
 \end{pmatrix} ~.
\end{align}
All our notations and conventions regarding osp(2$|$2) may be
found in appendix \ref{App:sl12}. In this parametrization, the WZNW
action becomes
\begin{align}
  S^{\tWZNW} (g)
   &=\ \frac{k}{2\pi} \int d^2 z \, \left[\,  \partial \varphi_1 \partial \varphi_1
   + \partial \phi_2 \bar \partial \phi_2 +   e^{- i \varphi_1 - \phi_2}
     \bar \partial \theta_1 \partial \bar \theta_2
 + e^{i \varphi_1 - \phi_2} \bar \partial \theta_2 \partial \bar \theta_1\,  \right] \\[3mm]
     &+ \frac{k}{2\pi} \int d^2 z \, e^{-2 \phi_2}
    \left( \partial \bar \gamma + \frac12 (
   \bar \theta_1 \partial \bar \theta_2
 +  \bar \theta_2 \partial \bar \theta_1) \right)
    \left( \bar \partial \gamma - \frac12 (\theta_1 \bar \partial \theta_2
 + \theta_2 \bar \partial \theta_1 ) \right) \nonumber ~.
\end{align}
Introducing new variables as before, the action is rewritten as
\begin{align}
  S^{\tWZNW} (g)
   &=\ \frac{1}{2\pi} \int d^2 z
 [ \partial \varphi_1 \bar \partial \varphi_1
   + \partial \phi_2 \bar \partial \phi_2  \\[2mm]
 &+ \beta \left( \bar \partial \gamma - \frac12 (
   \theta_1 \bar \partial  \theta_2
 +  \theta_2 \bar \partial \theta_1 ) \right)
 + \bar \beta \left( \partial \bar \gamma + \frac12 (
   \bar \theta_1 \partial \bar \theta_2
 +  \bar \theta_2 \partial \bar \theta_1) \right)
  - \frac{1}{k} \beta \bar \beta e^{ 2b \phi}
 \nonumber \\[2mm]
 &+ p_1 \bar \partial \theta_1 + \bar p_1 \partial \bar \theta_1
 + p_2 \bar \partial \theta_2 + \bar p_2 \partial \bar \theta_2
 - \frac{1}{k} p_1 \bar p_2 e^{ b ( i \varphi_1 + \phi_2 )}
 - \frac{1}{k} p_2 \bar p_1 e^{ b (- i \varphi_1 + \phi_2 )} ]
 \nonumber
\end{align}
with the parameter $b$ being related to the level $k$ of the
WZNW model through $b = 1/\sqrt{k - 1}$. Note that this relation
differs from the one we found in the case of  \osp.

In the first order formulation the $N$-point function of tachyon
vertex operators becomes
\begin{align}
 \langle \prod _{\nu=1}^N V_{j_\nu} (\mu_\nu | z_\nu) \rangle
  &=\   \int   {\cal D} \varphi_1 {\cal D} \phi_2
 {\cal D}^2 \beta {\cal D}^2 \gamma \prod_{a=1}^2 {\cal D}^2 \theta_a
  {\cal D}^2 p_a\,  e^{-S^{\tWZNW} [g]} \prod_{\nu=1}^{N}
  V_{j_\nu} (\mu_\nu | z _ \nu) ~.
\end{align}
In our discussion of the OSP(2$|$2) WZNW  model we shall restrict
ourselves to correlation functions of purely bosonic vertex operators,
\begin{align}
 V_{j_\nu} (\mu_\nu | z_\nu) \ = \ | \mu_\nu |^{2j_\nu + 2}
 e^{\mu_\nu \gamma - \bar \mu_\nu \bar \gamma}
  e^{2 b (j_\nu +1) \phi} ~.
\end{align}
It is not too difficult to work with the full vertex operators,
including exponentials of the fermionic fields. Since we have
seen above how such terms are taken into account, there is no
need to repeat all this for osp(2$|$2) now. Following \cite{HS}
we integrate out $\beta,\gamma$ to obtain
\begin{align}
  \langle \prod _{\nu=1}^N V_\nu (z_\nu) \rangle
  &=\  \delta^2 (\sum_{\nu = 1}^N \mu_\nu) 
  |\tilde  \Theta_N|^4\prod_{a=1}^2
 \int {\cal D} \varphi_a  {\cal D}^2 \theta_a {\cal D}^2 p_a \times
 \nonumber \\[2mm]  &\times\
   e^{-S [\varphi_a,\theta_a,p_a]}
   \prod_{\nu=1}^{N} e^{2 ( b(j_\nu+1) + \frac{1}{2b}) \varphi_2} (z_\nu)
    \prod_{j=1}^{N-2} e^{ - \frac{1}{b} \varphi_2 } (y_j) ~,
\end{align}
where the action is
\begin{align}
   S [\varphi_a,\theta_a,p_a]
   &=\  \frac{1}{2\pi} \int d^2 z [ \partial \varphi_1 \bar \partial \varphi_1
    +  \partial \varphi_2 \bar \partial \varphi_2
   + \frac{Q_2}{4} \sqrt g R \varphi_2  \\[2mm]
 &- \frac12 u {\cal B}
  ( \theta_1 \bar \partial \theta_2 + \theta_2 \bar \partial \theta_1)
 - \frac12 \bar u \bar {\cal B}
  (\bar \theta_1 \partial \bar \theta_2 + \bar \theta_2 \partial \bar \theta_1)
 + \frac{1}{k} e^{ 2 b \varphi_2}
 \nonumber \\[2mm]  &
 + p_1 \bar \partial \theta_1 + \bar p_1 \partial \bar \theta_1
 + p_2 \bar \partial \theta_2 + \bar p_2 \partial \bar \theta_2
 + \frac{i}{k} \frac{1}{|u{\cal B}|} ( p_1 \bar p_2
   e^{ b (i \varphi_1 + \varphi_2) } + p_2 \bar p_1
   e^{ b (-i \varphi_1 + \varphi_2) } )] \nonumber
\end{align}
with $Q_2 = 1/b$. The factor $\tilde \Theta_N$ has been introduced
previously in eq.\ \eqref{ttheta}. Now we follow the same steps
as in the case of \osp, i.e.\ we rescale and rotate all the
fermionic fields,
\begin{align}
 \chi_1 &:= \ i (\theta_1 \sqrt{u {\cal B}} -p_2 / \sqrt{u {\cal B}}  )~,
 &\chi_2 &:=\ i (\theta_2 \sqrt{u {\cal B}} - p_1 / \sqrt{u {\cal B}}) ~,
 &\psi_a &:=\  p_a / \sqrt{u {\cal B}}  ~.
\end{align}
Once more, it is not difficult to rewrite the action in terms
of the new fermionic fields in terms of $\chi_a,\psi_a$. For the
kinetic terms this is done with the help of the identity,
\begin{align}
&- \frac12 u {\cal B}
  ( \theta_1 \bar \partial \theta_2 + \theta_2 \bar \partial \theta_1)
 + \sum_a p_a \bar \partial \theta_a \\[2mm]
 &\qquad = \ \chi_1 \bar \partial \chi_2
  + \psi_1 \bar \partial \psi_2
  +  \sum_a \left(i \psi_a \chi_a - \frac12\partial  \ln
  \sqrt{u {\cal B}} \right)
  \bar \partial \ln \sqrt{u {\cal B}} ~.
  \nonumber
\end{align}
In order to spell out the vertex operators, we bosonize all four
fermionic fields,
\begin{align}
 \psi_a \pm i \chi_a \ = \ 2 \exp (\pm i Y_a)  ~.
\end{align}
Putting all this together, we arrive at the following formula for
correlation functions in the WZNW model,
\begin{align}
  \langle \prod _{\nu=1}^N V_{j_\nu} (\mu_\nu | z_\nu) \rangle
  &=\  \delta^2 (\sum_{\nu = 1}^N \mu_\nu)\, | \Theta^{(2)}_N|^2
  \, \prod_{a=1}^2
 \int {\cal D} \varphi_a  {\cal D}^2 \psi_a {\cal D}^2 \chi_a
   e^{-S [\phi_a,\psi_a,\chi_a]}  \times \\[2mm]
    &\times
   \prod_{\nu=1}^{N} e^{  \frac{i}{2 }\sum_a ( Y_a + \bar Y_a )
    + 2 ( b(j_\nu+1) + \frac{1}{2b}) \varphi_2} (z_\nu)
    \prod_{j=1}^{N-2} e^{ - \frac{i}{2 }\sum_a ( Y_a + \bar Y_a )
    - \frac{1}{b} \varphi_2} (y_j) ~, \nonumber
\end{align}
where the action $S$ is built from an ${\cal N}=2$ supersymmetric
Liouville theory for the fields $(\varphi_a,\psi_a,\bar \psi_a)$
and the theory of two free fermions $(\chi_a,\bar \chi_a)$,
\begin{align}
   S [\phi,\psi,\chi]
   &=\  \frac{1}{2\pi} \int d^2 z [
  \partial \varphi_1 \bar \partial \varphi_1  +
   \partial \varphi_2 \bar \partial \varphi_2
   + \frac{Q_2}{4} \sqrt g R \varphi_2
    + \frac{1}{k} e^{ 2 b \varphi_2}
    \\[2mm]
 & + \psi_1 \bar \partial \psi_2 + \bar \psi_1 \partial \bar \psi_2
  + \chi_1 \bar \partial \chi_2 + \bar \chi_1 \partial \bar \chi_2
 + \frac{i}{k}
 (\psi_1 \bar \psi_2 e^{ b (i \varphi_1 + \varphi_2) }
 + \psi_2 \bar \psi_1 e^{ b (-i \varphi_1 + \varphi_2)})] ~.
 \nonumber
\end{align}
The factor $\Theta^{(2)}_N$ is very similar to the corresponding 
function in the \OSP\ model, 
\begin{align}
  \Theta_N^{(2)} \ = \ u
 \prod_{\mu < \nu}^N (z_\mu - z_\nu)^{\frac{1}{2b^2}-\frac12}
 \prod_{i < j}^{N-2} (y_i - y_j)^{\frac{1}{2b^2}-\frac12}
 \prod_{\nu = 1}^N \prod_{i=1}^{N-2} 
(z_\nu - y_i)^{-\frac{1}{2b^2}+\frac12}  ~.
\end{align}
In conclusion, we have shown that correlators of tachyon vertex
operators in the OSP(2$|$2) WZNW model may be obtained from
correlation functions in ${\cal N}=2$ Liouville field theory. Since
we do not know much about the latter when ${\cal N} \geq 3$, we shall
not exploit this relation any further. Instead, we shall now
focus on the \OSP\ WZNW model.


\section{\OSP\ WZNW model - the particle limit}

Our main aim below is to exploit our relation \eqref{H2L} between
the \OSP\ WZNW model and ${\cal N}=1$ Liouville theory along with known
results for the latter in order to solve the former. Before going
into the full and cumbersome conformal field theory computations,
however, it is instructive to examine the particle limit. We
shall determine the minisuperspace wave functions in the first
subsection and then calculate particle analogues of the
2-point and 3-point functions.

\subsection{Particle wave functions on \OSP}

Let us first construct the wave functions for a particle that
moves freely on the supergroup \OSP. The mathematical problem
that needs to be solved is to determine all eigenfunctions of
Laplace operator. In our parametrization \eqref{paraosp}, the
generators of infinitesimal right translations on \OSP\ are
easily worked out,
\begin{align}\label{rreg}
 R_{E^+} & =\ \partial_{\gamma} \ ~, \quad \quad \quad
  R_{H} \ =\ - \frac12 \partial_\phi - \gamma \partial_\gamma
 - \frac12 \theta \partial_\theta ~, \nonumber \\[2mm]
  R_{E^-} & =\  e^{2 \phi} \partial_{\bar \gamma} -
   \gamma ^2 \partial_\gamma
 -  \gamma \partial_\phi - \gamma \theta \partial_\theta
 +  e^{\phi} \theta (\partial_{\bar \theta}
 - \bar \theta \partial_{\bar \gamma} ) ~,  \\[2mm]
 R_{F^+} & = \ \frac12 (\partial_\theta + \theta \partial_\gamma) ~,
  \quad R_{F^-}  =\ \frac12 e^{\phi} (\partial_{\bar \theta}
 -  \bar \theta \partial_{\bar \gamma} )
 - \frac12  \gamma (\partial_\theta + \theta \partial_\gamma)
 - \frac12 \theta \partial_\phi  ~. \nonumber 
\end{align}
We may now insert these expressions into the general formula for
the quadratic Casimir element and thereby derive the following
Laplacian on \OSP\
\begin{align}
 \Delta &=\  H H + \frac12 (E^+ E^- + E^- E^+ )-  (F^+ F^- -  F^- F^+ )
 \nonumber \\[2mm]
&=\  \frac14 \partial_\phi ^2 - \frac14 \partial_\phi +
   e^{2 \phi} \partial_\gamma \partial_{\bar \gamma} -
 \frac12 e^{\phi} ( \partial_\theta - \theta \partial_\gamma )
 (\partial_{\bar \theta} - \bar \theta \partial_{\bar \gamma} ) ~.
\end{align}
The construction of eigenfunction proceeds in several steps. To
begin with, we shall look for eigenfunctions $\Phi^j$ of the
following special form
\begin{align}
 \Phi^j(\gamma,\bar \gamma,\theta,\bar \theta,\phi)\  =
 \ e^{\mu \gamma - \bar \mu \bar \gamma}
  \, \phi^j_{\mu \bar \mu} (\theta,\bar \theta,\phi) ~.
\end{align}
Here, $j$ parametrizes the eigenvalue $\lambda =
(j+1)(j+\frac12)$ of the Laplacian. With our ansatz,
we have explicitly diagonalized
the operators $R_{E^+}$ and $L_{E^-} = \partial_{\bar \gamma}$. The
functions $\Phi^j$ give rise to eigenfunctions of the Laplacian,
provided that the factor  $\phi^j_{\mu \bar \mu}$ is an eigenfunction
of the operator
\begin{align}
 \Delta_\mu &=\  \frac14 \partial_\phi ^2 - \frac14 \partial_\phi -
  e^{2 \phi} \mu \bar \mu - \frac12
 e^{\phi}\, \Xi ~
&\text{where} \ \ \ \Xi_\mu &=\  ( \partial_\theta - \theta \mu )
 ( \partial_{\bar \theta} + \bar \theta \bar \mu  ) ~.
\end{align}
Here and in the following we shall assume that $|\mu| ^2 \equiv
\mu \bar \mu > 0$. Our second step now is to diagonalize the
operator $\Xi_\mu $ on the 4-dimensional Grassmann algebra that is
generated from $\theta$ and $\bar \theta$. We can easily find two
eigenfunctions $S^\pm_\mu, T^\pm_\mu$ for each of the eigenvalues
$\xi_\mu = \pm |\mu|$,
\begin{align}\label{Spmmu}
 S^{+}_\mu (\theta , \bar \theta) &=\  \frac{1}{\sqrt2 } 
    (1 -  |\mu | \theta \bar \theta)  ~,
 &S^{-}_\mu(\theta , \bar \theta)  &=\ \frac{1}{\sqrt2 } 
    (1 + |\mu | \theta \bar \theta ) ~,
 \\[2mm] \label{Tpmmu}
 T^{+}_\mu(\theta , \bar \theta)  &=\ \frac{1}{\sqrt2}
  (\sqrt{\mu}  \theta - \sqrt{\bar \mu}  \bar \theta )~,
 &T^{-}_\mu(\theta , \bar \theta)  &=\  \frac{1}{\sqrt2 }
  (\sqrt{\mu}  \theta + \sqrt{\bar \mu}  \bar \theta )~.
\end{align}
Having solved the eigenvalue problem for the operator $\Xi_\mu$,
we can now split off the dependence of the eigenfunctions
$\phi^j_{\mu\bar \mu}$ on fermionic coordinates through the
following ansatz
\begin{align}
\phi^{j,\pm}_{\mu \bar \mu} (\theta, \bar \theta , \phi)
 &=\  S^{\pm}_\mu (\theta , \bar \theta) U_p^{\pm} (\rho) ~,
&\psi^{j,\pm}_{\mu \bar \mu} (\theta, \bar \theta , \phi) &=
T^{\pm}_\mu (\theta , \bar \theta) U_p^{\pm} (\rho)\ \ .
\end{align}
Here we have introduced the letter $\psi$ instead of $\phi$ for
odd eigenfunctions. Furthermore, the variable $\rho$ on the right
hand side is defined by $\rho = 2 |\mu| e^\phi$. A short computation
shows that the functions $U^\pm_p$ possess no explicit $\mu$
dependence any more. There remains a dependence on the parameter
$j = -3/4 + ip/2$ which we display explicitly through the
subscript $p$. In any case, the functions $\phi^{j,\pm}_{\mu\bar \mu}$
and $\psi^{j,\pm}_{\mu\bar \mu}$ give rise to eigenfunctions of
our Laplacian provided that $U^\pm_p(\rho)$ satisfy the second order
differential equation
\begin{align}
  (\partial_\rho^2 - 1 \mp \frac1\rho + \frac{1/4 + p^2}{\rho^2}) U^{\pm}_p(\rho)
\ = \ 0 ~.
\end{align}
Solutions to these equations are
\begin{align}\label{Upmp}
 U^{\pm}_p (\rho) \ = \ \frac{ {\cal C}^{i p}}{\Gamma(\frac12 - i p ) }
  \ (\rho K_{\frac12 + i p} (\rho) \mp \rho K_{\frac12 - i p} (\rho) ) ~.
\end{align}
In order to single out some unique solution we demanded regularity
at $|\rho| \to \infty$. The normalization constant ${\cal C}$ remains
undetermined for now.  \smallskip

In summary we have obtained a basis of eigenfunctions of the
Laplacian $\Delta$ in \OSP. Grassmann even functions with
eigenvalue $\lambda = \lambda_p = (j+1)(j+\frac12) =
-\frac{1}{4}(\frac{1}{4} + p^2)$ take the form
\begin{equation}
\Phi^{j,\pm}_{\mu \bar \mu}(\gamma,\bar
\gamma,\theta,\bar\theta,\phi) \ = \ e^{\mu \gamma - \bar \mu \bar
\gamma} \, S^{\pm}_{\mu}(\theta,\bar \theta) \, U_p^\pm (2|\mu|
e^\phi)\label{eef}
\end{equation}
where $\mu,\bar \mu$ run through the complex plane. The functions
$U^\pm_p$ and $S^\pm_\mu$ have been defined in eqs.\ \eqref{Upmp}
and \eqref{Spmmu}, respectively. A similar formula with
$T^\pm_\mu$ instead of $S^\pm_\mu$ holds for Grassmann odd
eigenfunctions $\Psi^{j,\pm}_{\mu \bar \mu}$.
\medskip

Let us conclude this subsection by computing the analogue of the
minisuperspace 2-point function for Grassmann even functions in
the \OSP\ model.\footnote{2-point functions for Grassmann odd
functions $\Psi^{j,\epsilon}_{\mu,\bar \mu}$ can be computed in
the same way. We will not discuss those here, neither in the
particle model nor in the field theory.} This requires integrating
the product of two functions \eqref{eef} over the supergroup \OSP.
Using the Haar measure $[ d g ] = e^{- \phi} d \phi d \bar \gamma d
\gamma d \bar \theta d \theta$ on \OSP\ we obtain
\begin{align}\nonumber
\langle\, \Phi^{j_1,\epsilon_1}_{\mu_1,\bar \mu_1}
  \, \Phi^{j_2,\epsilon_2}_{\mu_2,\bar \mu_2}\, \rangle_0 & := \
 \int [ d g ] \, \Phi^{j_1,\epsilon_1}_{\mu_1,\bar \mu_1} (\gamma,\bar \gamma,\theta, \bar \theta,\phi)
\phi^{j_2,\epsilon_2}_{\mu_2 \bar \mu_2}
(\gamma,\bar\gamma ,\theta, \bar \theta,\phi ) \\[2mm]
\label{2ptmue} & = \  - |\mu_2| \epsilon_2 \delta_{\epsilon_1 ,
\epsilon_2}
   \delta^2 (\mu_1 + \mu_2 ) \int \frac{d x}{x^2}
   U^{\pm}_{p_1} ( 2 |\mu_1| x ) U^{\pm}_{p_2} ( 2 |\mu_2| x )\\[4mm]
& =\ - 2 \pi |\mu_2|^2 \epsilon_2 \delta_{\epsilon_1 , \epsilon_2}
   \delta^2 (\mu_1 + \mu_2 )
  \left(\delta ( p_1 + p_2 ) + {\cal R}^{\epsilon_2} (p_2) \delta (p_1 - p_2)\right) ~.
\nonumber
\end{align}
To get from the second to the third line we have utilized a
formula for integrals of Bessel functions, see appendix \ref{int}.
The reflection amplitude ${\cal R}$ in the last line is given by
\begin{align}
 {\cal R}^{\pm}(p) \ = \ \mp {\cal C}^{2 i p} \frac{\Gamma ( \frac{1}{2} + i p )}
   {\Gamma (\frac{1}{2} - i p)}
 ~. \label{mini2}
\end{align}
We shall later compare this answer with the outcome of a full
fledged conformal field theory computation of the 2-point
function. We could now start to analyze 3-point functions but
before we do so, we would like to talk about another basis in the
space of functions on \OSP.

\subsection{Wave functions - another basis}

In the last subsection, we constructed a basis for the space of
functions on \OSP. This basis is very convenient for describing
the duality between the \OSP\ WZNW model and ${\cal N}=1$ Liouville field
theory. When it comes to writing down explicit formulas for
correlation functions, on the other hand, there exists another,
preferable choice. Recall that all correlators contain an \osp\
invariant tensor that is determined by symmetries alone, along
with the structure constants which contain all dynamical
information. While the latter are the same in every basis, the
former depend very much on our choices. We shall now present a new
basis in which the \osp\ invariant tensors take a particularly
simple form.
\smallskip

The transformation from the old basis \eqref{eef} to the new can
be thought of as a Fourier/Bessel transform in $\mu,\bar \mu$ and
their fermionic counterparts. Let us perform the fermionic
transformation first. This amounts to defining new functions
$\Phi^j_{\mu}$ by
\begin{align} \label{phimuxi}
 \Phi^j_{\mu} (\xi|\gamma,\theta,\phi)
  & :=\ \frac{1}{|\mu|} \sum_{\epsilon=\pm} \left(S^{\epsilon}_\mu(\xi)
  \Phi^{j,\epsilon}_{\mu}(\gamma,\theta,\phi) +
  T^{\epsilon}_\mu(\xi)
  \Psi^{j,\epsilon}_{\mu}(\gamma,\theta,\phi)\right) ~.
\end{align}
Here and throughout the rest of this section we shall suppress
spelling out the dependence on the bared quantities such as $\bar
\xi, \bar \theta, \bar \gamma, \bar \mu$. Following \cite{Ribault:2005wp}, we
also transform from the variables $\mu,\bar \mu$ to new variables
$x,\bar x$,
\begin{align}
 \Phi^j (x,\xi|\gamma,\theta,\phi)
  \ := \  \frac{1}{4\pi} |\mu|^{-2j - 2} \int d^2 \mu
\,   \Phi^j_\mu (\xi|\gamma,\theta,\phi)\ e^{\bar\mu \bar x - \mu  x}~.
\label{xbasis}
\end{align}
We shall often refer to $\Phi^j(x,\xi|\gamma,\theta,\phi)$ as
eigenfunctions in the {\em $(x,\xi)$-basis}. For given label $j$,
the generators \eqref{rreg} of right translations may be expressed
through their action on the auxiliary variables $x,\bar x, \xi,
\bar \xi$ as follows,
\begin{align}
 &r_{E^+} = \ \partial_{x} ~, \qquad
 r_{H}\  = \  - x \partial_x - \frac12 \xi \partial_\xi + j+\frac12~, \nonumber
 \\[2mm]
 &r_{E^-} \ =\  - x ^2 \partial_x  - x \xi \partial_\xi + x (2j + 1) ~,
 \label{rlist} \\[2mm]
 &r_{F^+}\  = \ \frac12 (\partial_\xi + \xi \partial_x) ~, \qquad
 r_{F^-}\  =\
 - \frac12 x (\partial_\xi + \xi \partial_x)
 + \xi (j+\frac12)  ~. \nonumber
\end{align}
Given these expressions it is rather straightforward to verify
that \osp-invariance fixes the minisuperspace 2-point function to
be of the form
\begin{align}\label{2ptxxi}
 \langle \Phi^{j_1} (x_1,\xi_1) \Phi^{j_2} (x_2,\xi_2) \rangle
 = \frac{\pi}{4}
   \left[\delta ^2 (X_{12} )  \delta(j_1 + j_2 + 3/2 )
  + \frac{B^\Hrep(j_2)}{| X_{12} |^{ -4 j_2 - 2}} \delta(j_1 - j_2 )\right]~,
\end{align}
up to an overall normalization which we have fixed such that the
coefficient of $\delta$ functions in the first term is $\pi/4$. In
the first term we have also used the shorthand
\begin{align}
\delta ^2 (X_{12} ) \ = \ \delta^2 (x_{12})
 (\xi_1 - \xi_2)( \bar \xi_1 - \bar \xi_2 ) ~.
\end{align}
Furthermore, we employed the notation $X_{ij} = x_i - x_j - \xi_i
\xi_j$ that will appear frequently throughout the rest of this
note. The non-trivial structure constant $B^\Hrep(j)$ is not determined
by symmetry. It may be calculated by explicitly performing the
integral over the group manifold. Here, we shall follow a slightly
different route. Our aim is to relate the two formulas
\eqref{2ptmue} and \eqref{2ptxxi} for the 2-point functions. We
can then read off $B^\Hrep(j)$ from our formula \eqref{mini2} for the
reflection amplitude ${\cal R}^{\pm}(p)$. With the help of some
integral formulas that are spelled out in appendix \ref{int},
one may show that
\begin{align}
 \langle \Phi^{j_1,\epsilon_1}_{\mu_1}
 \Phi^{j_2,\epsilon_2}_{\mu_2}\rangle_0
  = - \pi |\mu_2|^2 \epsilon_2 
  \delta_{\epsilon_1,\epsilon_2} \delta^2(\mu_1+\mu_2)
  [ \delta(j_1+j_2 + \tfrac{3}{2}) - \epsilon_2 B^\Hrep(j_2) \pi \gamma (2j_2 + 1)
   \delta( j_1 - j_2 ) ] \nonumber
\end{align}
where $\gamma (x) = \Gamma(x)/\Gamma(1-x)$. Indeed, this outcome 
is fully consistent with our previous formula \eqref{2ptmue}. The 
comparison also allows us to determine the structure constant 
$B^\Hrep$,
\begin{align}  B^\Hrep(j) & = \
  - \frac{\epsilon  {\cal R}^{\epsilon}(p)  } 
 {\pi \gamma ( 2 j + 1)} \ = \ - \frac{1}{\pi} \ {\cal C}^{2ip}\ \ ,
\label{miniB} \end{align}
where $j$ and $p$ are related by $j = -3/4 + ip/2$, as usual. By
comparing the
two expressions \eqref{2ptxxi} and \eqref{2ptmue} for the 2-point
function we have confirmed that the expression \eqref{2ptmue} is
consistent with \osp\ invariance and we have determined the
structure constant $B^\Hrep(j)$ that was introduced in eq.\
\eqref{2ptxxi}.

\subsection{The minisuperspace 3-point function}

We are now prepared to move on to the analogue of the 3-point
function in the particle model. Once more, the symmetry under
\osp\ transformation fixed the 3-point function up to two
structure constants. In the $(x,\xi)$-basis, it reads
\begin{align}\label{3ptxxi}
 &\langle \Phi^{j_1} (x_1,\xi_1) \Phi^{j_2} (x_2,\xi_2)
          \Phi^{j_3} (x_3,\xi_3) \rangle_0 \ =\
   \frac{C^\Hrep (j_1,j_2,j_3) +  \tilde  C^\Hrep (j_1,j_2,j_3) \eta \bar
   \eta}
   {| X_{12} |^{-2 j_{12}-1}
            | X_{23} |^{-2 j_{23} -1} | X_{31} |^{-2
            j_{31}-1}}~,
\end{align}
where $j_{12} = j_1 + j_2 - j_3$ etc.\ and $X_{ij} = x_i - x_j -
\xi_i\xi_j$, as before. Furthermore, the so-called super-projective
3-point invariants $\eta,\bar \eta$ are
\begin{align}\label{eta}
 \eta\ = \ (x_{12} x_{23} x_{31})^{-\frac12} (x_{23} \xi_1 + x_{31} \xi_2 +
  x_{12} \xi_3 - \tfrac12 \xi_1 \xi_2 \xi_3 ) 
\end{align}
and similarly for $\bar \eta$.%
\footnote{Since the \osp\ superalgebra has super dimension $3|2$,
we can fix three bosonic parameters $x$ and two fermionic parameters $\xi$.
Therefore, the three point function still depends on a fermionic invariant 
$\eta$.} The coefficients $C^\Hrep,\tilde C^\Hrep$ cannot be determined 
from \osp\ invariance. Instead, they require to 
perform the full integral over the 
supergroup manifold. Since we do not need the results, we are not going 
to compute the minisuperspace coefficients $C^\Hrep$ and $\tilde C^\Hrep$ 
explicitly. Determining their field theoretic analogue is one of 
the main issues in the next section. 
\smallskip 

Before we conclude this section, we would like to deduce from eq.\ 
\eqref{3ptxxi} the 3-point function for the even part 
$\Phi^\text{B}$ of 
our functions \eqref{phimuxi} in the mixed basis,
\begin{align} \label{phiBmuxi}
\Phi^{\text{B},j}_{\mu} (\xi|\gamma,\theta,\phi) & := \ \left(
\Phi^{j}_{\mu} (\xi|\gamma,\theta,\phi)\right)^{\text{even}} \ =\
 \frac{1}{|\mu|} \sum_{\epsilon=\pm} S^{\epsilon}_\mu(\xi)
  \Phi^{j,\epsilon}_{\mu }(\gamma,\theta,\phi)\ \ .
\end{align}
The field theoretic analogues of these functions shall feature in
the next section when we discuss correlators of the \OSP\ WZNW
model. A short and straightforward computation shows that
\begin{align}
  \langle \prod_{i=1}^3 \Phi^{\text{B},j_i}_{\mu_i,\bar \mu_i}
(\xi_i) \rangle_0
 & = \ C^\Hrep (j_1,j_2,j_3) \left( D_{0} [j_i , \mu_i ]
  + \frac12 \sum_{a,b,c=1}^3 \epsilon_{abc} D_{bc}   [j_i , \mu_i ]
  \xi_b \bar \xi_b \xi_c \bar \xi_c \right)
  \nonumber \\[2mm]
 &\hspace*{0cm} + \, \tilde  C^\Hrep (j_1,j_2,j_3) \left( \sum_{a=1}^3 D_{a}
   [j_i , \mu_i ] \xi_a \bar \xi_a
  + D_{123} [j_i , \mu_i ]   \xi_1 \bar \xi_1 \xi_2 \bar \xi_2 \xi_3 \bar \xi_3
    \right) \label{3ptmuxi}
\end{align}
where $C^\Hrep$ and $\tilde C^\Hrep$ are the same as before and 
$\epsilon_{123} = 1$ and so on.
Various group theoretic factors are given by
\begin{align} \label{CGC3}
 &D_{0} [j_i, \mu_i ] \, =\ [ \prod_{i=1}^3|\mu_i|^{-1} ]
D \left[ \begin{smallmatrix}
j_1 + \frac12 &j_2  + \frac12& j_3  + \frac12 \\
\mu_1 & \mu_2 & \mu_3
\end{smallmatrix} \right]
~, 
\nonumber
 \\[2mm]  
&D_{123} [j_i, \mu_i ] \, = \ - (j + 2)^2 D \left[
\begin{smallmatrix}
j_1 &j_2& j_3 \\
\mu_1 & \mu_2 & \mu_3
\end{smallmatrix}\right]
~,  \\[3mm] 
 &D_{1} [j_i, \mu_i ] \, =\, |\mu_2|^{-1} |\mu_3|^{-1}
D \left[ \begin{smallmatrix}
j_1  &j_2  + \frac12& j_3  + \frac12 \\
\mu_1 & \mu_2 & \mu_3
\end{smallmatrix}\right]
~,
\nonumber  \\[3mm] 
&D_{12} [j_i \mu_i ] \, = \, - |\mu_3|^{-1}(j_{12} +
{\textstyle \frac{1}{2}} )^2 D \left[ \begin{smallmatrix}
j_1 &j_2& j_3 + \frac12 \\
\mu_1 & \mu_2 & \mu_3
\end{smallmatrix}\right] ~,
\nonumber 
\end{align}
and those with index permutations.
On the right hand side there appears the single new function $D$
that is defined in \eqref{CGC2}.
As we anticipated, the final expression for the 3-point function
in the $\mu$-basis turns out to be rather involved simply because
the group theoretic contributions to the structure constants are
rather complicated.


\section{Solution of the \OSP\ WZNW model}

In this section we compute the structure constants of
\OSP\ WZNW model. As we discussed in the previous section,
the 2- and 3-point functions are almost fixed by \osp\
symmetry, up to three coefficients $B_b^\Hrep$, $C_b^\Hrep$, $\tilde
C_b^\Hrep$ that remain undetermined and turn out to acquire field
theoretic modifications. In order to fix these coefficients we
utilize the results of section \ref{secSRT}, where we have derived
the relation between correlators of \OSP\ model and ${\cal
N}=1$ super Liouville theory.

\subsection{The WZNW-Liouville correspondence revisited}

When we discussed the correspondence between the \OSP\ WZNW model and
${\cal N}=1$ Liouville theory we worked with vertex operators
$V^{\epsilon,\bar \epsilon}_{j}(\mu|z)$ containing both bosonic
and fermionic components. In our computations here we shall
restrict ourselves to correlators involving purely bosonic fields since they
are enough to fix the unknown functions. The bosonic component of
$V^{\epsilon,\bar \epsilon}_{j}(\mu|z)$ is a field theoretic
analogue of the function \eqref{eef} on $\OSP$, i.e.\
\begin{align} \label{Veven}
V^{\epsilon}_j(\mu|z) & := 
 \ \frac{1}{\sqrt2} \left(V^{\epsilon',\bar
\epsilon'}_j(\mu|z)\right)^{\text{even}} \ = \ 
\frac{1}{\sqrt2} \left(1 - \epsilon
|\mu| \theta \bar \theta\right) \ |\mu|^{2j+2} e^{\mu\gamma - \bar
\mu \bar \gamma} e^{2b(j+1) \phi}
\end{align}
where $\epsilon = - \epsilon' \bar \epsilon'$.  Note that
$V^{\epsilon}_j(\mu|z)$ are indeed modeled after the functions
\eqref{eef}, i.e.\
$$
V^{\pm}_j(\mu|z) \ = \ S^{\pm}_\mu(\theta,\bar \theta)\, e^{\mu
\gamma - \bar \mu \bar \gamma}\,  |\mu|^{2j+1} e^{2(j+1) \phi}
$$
where $S^\pm_\mu$ is the same as in eq.\ \eqref{Spmmu}. Only
now the symbol $\theta = \theta(z)$ denotes a fermionic field on
the world-sheet and similarly for $\bar \theta$. Under
the change of variables described in section 2, the field
$S^{\pm}_\mu$ becomes
$$ S^{\pm}_\mu \ = \ \frac{1}{\sqrt{2}} \left( 1 \pm i \theta'\bar
\theta'\right) \ = \ \frac{1}{\sqrt{2}} \left( 1 \pm  \frac{i}{2}
(\psi-i\chi)( \bar \psi + i \bar \chi)\right) \ = \
\frac{1}{\sqrt{2}} \left( 1 \pm  e^{-iY + i \bar Y}\right) ~.$$
With this preparation we can now compute
correlation functions of the fields $V^\epsilon_j(\mu|z)$ in the
\OSP\ WZNW model through our relation \eqref{H2L} with ${\cal N}=1$
Liouville field theory,
\begin{align} \label{sRTeven}
  \langle \prod _{\nu=1}^N V^{\epsilon_\nu}_{j_\nu}
  (\mu_\nu | z_\nu) \rangle
  &= 
    \delta^2 (\sum_{\nu = 1}^N \mu_\nu) | \Theta_N^{(1)}|^2
  \langle \prod _{\nu=1}^N S^{\epsilon_\nu} e^{
      \alpha_\nu \varphi } (z_\nu)
    \prod_{j=1}^{N-2} e^{- \frac{i}{2} {\cal Y}}
  e^{- \frac{1}{2b} \varphi} (y_j)  \rangle^L
\end{align}
with $b=1/\sqrt{2k-3}$ and $\alpha_\nu = 2 b(j_\nu + 1) + 1/2b$.
The index $\ ^L$ on the right hand side reminds us that the
correlator is to be computed in the product of super Liouville
theory with a free fermion theory. Here we have defined
\begin{align}
S^+ = \sqrt2 \cos  \tfrac{\cal Y}{2} ~, \qquad
S^- =  \sqrt2 i \sin \tfrac{\cal Y}{2} ~, \qquad
 {\cal Y} = Y - \bar Y ~.
\end{align}
Since the fields $S^{\pm}$ include both the fermionic field $\chi$
of the free fermion theory and the fermion $\psi$ of ${\cal N}=1$
Liouville theory, it is not straightforward to apply the results
of ${\cal N}=1$ super Liouville field theory. In order to do so we
utilize the well-known construction of $S^\pm$ through
spin fields of the real fermions (see e.g.\ \cite{df}),  
\begin{align}\label{ising}
  &\langle \prod_{i=1}^{2m} S^+ (z_i)
  \prod_{j=2m+1}^{2n}\!\!\! S^- (z_{j}) \rangle \ = 
  \\ \nonumber & \qquad \ = \ (-1)^{n-m}
  \langle \prod_{i=1}^{2m} \Sigma^+_\chi (z_i) \prod_{j=2m+1}^{2n}\!\!\! \Sigma^-_\chi
  (z_{j})\ \rangle  \langle \prod_{i=1}^{2m} \Sigma^+_\psi (z_i)
  \prod_{j=2m+1}^{2n}\!\!\! \Sigma^-_\psi (z_{j})
 \rangle  ~,  
\end{align}
where $\Sigma^\pm_\chi$ and $\Sigma^\pm_\psi$ are spin fields for
the real fermions $\chi$ and $\psi$, respectively.

\subsection{Computation of 2-point functions}

In order to practice using our relation \eqref{sRTeven}, we want
to compute the 2-point function of \OSP\ WZNW model. This case is
rather simple since no extra degenerate fields are to be inserted.
With eq. \eqref{sRTeven} and eq. \eqref{ising} we have
\begin{align}
&  \langle V^{\epsilon_1}_{j_1} (\mu_1|z_1) V^{\epsilon_2}_{j_2}
(\mu_2|z_2) \rangle  \ = \nonumber \\[2mm]
  & \quad \quad \quad = 
    \, \delta^2 (\mu_1 + \mu_2) |u|^2 |z_{12}|^{\frac{1}{2b^2}-\frac12}
     \langle S^{\epsilon_1} e^{\alpha_1 \varphi} (z_1)
             S^{\epsilon_2} e^{\alpha_2 \varphi} (z_2)
      \rangle ^L   \\[2mm]
  & \quad \quad \quad =\ -i e^{\frac{\pi i}{4}(\epsilon_1 + \epsilon_2)}
  \, \delta^2 (\mu_1 + \mu_2)\,  |u|^2 |z_{12}|^{\frac{1}{2b^2}-\frac12}
  \,   \langle \Sigma_\chi^{\epsilon_1} (z_1)
       \Sigma_\chi^{\epsilon_2} (z_2) \rangle
\ \langle \Sigma_{\alpha_1}^{\epsilon_1} (z_1)
         \Sigma_{\alpha_2}^{\epsilon_2} (z_2) \rangle  ~,\nonumber
\end{align}
where $\Sigma^{\pm}_{\alpha}$ are spin fields in ${\cal N}=1$ Liouville
theory, see eq.\ \eqref{spin} for a definition. Inserting the two
point function of spin fields in the free fermion theory,
\begin{align}
 \langle \Sigma_\chi^{\epsilon_1} (z_1) \Sigma_\chi^{\epsilon_2} (z_2 ) \rangle
  = \delta_{\epsilon_1 , \epsilon_2} | z_{12} |^{- \frac14} ~,
\end{align}
along with the corresponding formula for the 2-point function of
$\Sigma^\pm_\alpha$ in super Liouville theory, see eq.\
\eqref{L2pt}, the 2-point function of \OSP\ WZNW model can be
evaluated as
\begin{align}
&\langle V^{\epsilon_1}_{j_1} (\mu_1|z_1) V^{\epsilon_2}_{j_2}
(\mu_2|z_2) \rangle \,  = \, \delta^{\epsilon_1}_{\epsilon_2}
  \delta^2 (\mu_1 + \mu_2)\, \frac{\pi|\mu_2|^2\epsilon_2}{
   b|z_{12}|^{4 \Delta^\Hrep_{j_2}}}
 [\delta (j_1 + j_2 + \tfrac32)
  - \epsilon_2 \delta (j_1 - j_2 )
 D^L_R (\alpha_2) ]  \nonumber
\end{align}
with $\Delta^\Hrep_j = - 2 b^2 (j+1)(j+ \frac12)$. An explicit formula
for the structure functions $D^L_R$ of Liouville theory may be found
in eq.\ \eqref{refR}. By comparing our result with the general form
of the 2-point invariant \eqref{2ptxxi} we read off that
\begin{align}
 B^\Hrep_b(j)\  =\ 
  \frac{D^L_R (2b(j+1) + \frac{1}{2b})}{\pi \gamma (2 j + 1)}
   = - \frac{1}{\pi} \left(\frac{2 k b^2}{i \gamma (\tfrac{b^2 + 1}{2})}
   \right)^{4j+3}
     \frac{\Gamma(\frac12 + b^2 ( 2j + \frac32 ) )}
       {\Gamma(\frac12 - b^2 ( 2j + \frac32 ) )} ~.
\end{align}
In the limit $b \rightarrow 0$ we recover the result \eqref{miniB}
of the particle model.

\subsection{Computation of 3-point functions}

Our aim now is to determine the structure constants of the 3-point
function in the \OSP\ WZNW model from the correspondence with
${\cal N}=1$ Liouville theory. To this end we compute the 3-point
function of three bosonic vertex operators \eqref{Veven} using the
formula \eqref{sRTeven}.
\begin{align}
 \langle V^{\epsilon_1}_{j_1} (\mu_1|z_1) V^{\epsilon_2}_{j_2} (\mu_2|z_2)
   V^{\epsilon_3}_{j_3} (\mu_3|z_3) \rangle & = \
    \delta^2 (\mu_1 + \mu_2 + \mu_3) | \Theta_3^{(1)}|^2 \ 
    \ \times \
    \\[2mm]
  & \hspace*{-3cm} \times \ \langle \tfrac{1}{\sqrt2}(S^+ + S^-) 
  e^{-\frac{1}{2b}\varphi} (y)
    S^{\epsilon_1} e^{\alpha_1 \varphi}(z_1)
    S^{\epsilon_2} e^{\alpha_2 \varphi} (z_2)
    S^{\epsilon_3} e^{\alpha_3 \varphi}(z_3)
     \rangle ^L ~. \nonumber
\end{align}
Here, we use the same notations as in eq.\ \eqref{sRTeven} before.
Note that the computation of a 3-point function on the \OSP\ WZNW
model requires one additional insertion of a degenerate Liouville
field in the correlator on the right hand side. This field is
inserted at
\begin{align}
y &=\ - \frac{1}{u} (\mu_1 z_2 z_3 + \mu_2 z_3 z_1 + \mu_3 z_1
z_2) ~,
\end{align}
where the parameter $u$ is given by $u = \sum_{i=1}^3 \mu_i z_i$.
Furthermore, for $N=3$ the twist factor $|\Theta_3^{(1)}|^2$ 
defined in eq.\ \eqref{theta1} simplifies
as follows
\begin{align}
|\Theta_3^{(1)}|^2 \ = \ |u|^{\frac{3}{2b^2} + \frac{1}{2}}
  \prod_{i < j}|z_{ij}|^{-\frac{1}{2b^2}+ \frac12}
  \prod_{i=1}^3 | \mu_i|^{-\frac{1}{2b^2}+ \frac12} ~.
\end{align}
As in the case of the 2-point function we can express the
fields $S^\pm$ in terms of twist fields for the two fermions using
the formula \eqref{ising}. The correlator of four twist fields in
a free fermion model is known from the work of Belavin, Polyakov
and Zamolodchikov \cite{BPZ},
\begin{align}
 & \langle \Sigma_\chi^{\epsilon_0} (z_0) \Sigma_\chi^{\epsilon_1} (z_1)
   \Sigma_\chi^{\epsilon_2} (z_2) \Sigma_\chi^{\epsilon_3} (z_3) \rangle
 \ = \  | z_{03}|^{-\frac14} |z_{12}|^{-\frac14}
  {\cal I}^{\epsilon_0\epsilon_1\epsilon_2\epsilon_3} (z) ~.
\end{align}
Here $z \ (\equiv (z_{01}z_{23})/(z_{03}z_{21}))
 = 1 + \mu_2/\mu_3$ is the conformally invariant cross
ratio of the points $z_0 = y$ and $z_i, i=1,2,3$. The functions
${\cal I}^{\epsilon_0\epsilon_1\epsilon_2\epsilon_3} (z)$ are given by
\begin{align} 
  & {\cal I}^{\pm\pm\pm\pm} (z) = I_0 (z) I_0 (\bar z) 
  + I_{\frac12} (z) I_{\frac12} (\bar z) ~, \qquad
   {\cal I}^{\pm\pm\mp\mp} (z) =  I_0 (z) I_0 (\bar z) 
  - I_{\frac12} (z) I_{\frac12} (\bar z) ~, \nonumber \\
  & {\cal I}^{\pm\mp\pm\mp} (z) = \pm \left[
  I_0 (z) I_{\frac12 } (\bar z) + I_{\frac12} (z) I_0 (\bar z) \right]
 ~,  ~
   {\cal I}^{\pm\mp\mp\pm} (z) = i \left[ 
I_0 (z) I_{\frac12} (\bar z)  - I_{\frac12} (z) I_0 (\bar z) \right] 
 ~,  \nonumber 
\end{align}
with
\begin{align}
 I_0 (z) &= (z(1-z))^{-\frac18} 
   F(\tfrac{1}{4},-\tfrac{1}{4},\tfrac{1}{2}, z) ~,
 & I_{\frac12}(z) &= \tfrac12 (z(1-z))^{\frac38}
  F(\tfrac{5}{4},\tfrac{3}{4},\tfrac{3}{2}, z)~.
\end{align}
As for the contribution from Liouville theory, all relevant
formulas are listed in appendix \ref{Liouville}. The relevant
4-point function \eqref{L4pt} was constructed in \cite{FH}. It
involves a new function
${\cal H}^{\epsilon_0\epsilon_1\epsilon_2\epsilon_3}$ that we spell out
explicitly in eq.\ \eqref{feeee}. Putting all these pieces
together we obtain
\begin{align} \label{3PTmue}
 &\langle V^{\epsilon_1}_{j_1} (\mu_1|z_1)
 V^{\epsilon_2}_{j_2} (\mu_2|z_2) V^{\epsilon_3}_{j_3} (\mu_3|z_3) \rangle
 \ = \ \delta^2 (\mu_1 + \mu_2 + \mu_3) \
 \prod_{i<j} |z_{ij} |^{- 2 \Delta^{\Hrep}_{ij}}
    \  \times \\[2mm]& \qquad \times
     |\mu_1|^{-\frac{1}{2b^2}+ \frac12 }
     |\mu_2|^{-\frac{1}{2b^2}+ \frac12 }
     |\mu_3|^{\frac{1}{b^2} + 1 }
 \frac{e^{\pi i(4 - \epsilon - \sum \epsilon_\nu)/4}}{\sqrt2} \
 {\cal I}^{\epsilon\epsilon_1\epsilon_2\epsilon_3}(1+ \tfrac{\mu_2}{\mu_3})
  {\cal H}^{\epsilon\epsilon_1\epsilon_2\epsilon_3}(1+ \tfrac{\mu_2}{\mu_3})
  \nonumber ~,
\end{align}
where $\epsilon = \epsilon_1 \epsilon_2 \epsilon_3$. In principle we
have thereby completed our computation of the 3-point function in the
\OSP\ WZNW model. Of course, in its present form the answer is not
very illuminating, in particular when compared with the relatively
simple form of the 3-point function we anticipated in eq.\
\eqref{3ptx} of the introduction. The reason our formula \eqref{3PTmue}
looks somewhat unfamiliar was discussed in detail in section 3.3: It is
the transformation from the $x$ to the $\mu$ basis that turns the rather
simple looking formulas \eqref{3ptxxi} or \eqref{3ptx} into the bulky
expression of eqs.\ \eqref{3ptmuxi} or \eqref{3PTmue}. Our final task
is therefore to perform the transformation from eq.\ \eqref{3PTmue}
to \eqref{3ptx}. We shall not discuss this in full detail but simply
look at two of the terms in eq.\ \eqref{3ptmuxi} which suffice to
read off the structure functions $C_b^\Hrep$ and $\tilde C_b^\Hrep$.
\smallskip

Let us begin with the coefficient $\tilde C^\Hrep_b$ in eq.\ \eqref{3ptx}.
Comparison with our minisuperspace formula \eqref{3ptmuxi} shows that
$\tilde C^\Hrep$ appears in the coefficient of the term with the maximal
number of Grassmann variables. In fact, the coefficient of this term
is a product of $\tilde C^\Hrep$ with the group theoretic factor $D_{123}$.
In order to compare with our field theoretic outcome, we switch from
the $(\mu,\epsilon)$ basis to the mixed basis involving $\mu$ and $\xi$,
i.e.\ we rewrite the correlation function \eqref{3PTmue} in terms of
the fields
\begin{align}
 V^{\text{B}}_{j} (\mu,\xi|z)
  \ :=\   \left(V_{j} (\mu,\xi|z)\right)^{\text{even}} \ = \
   \frac{1}{|\mu|} \sum_{\epsilon = \pm} \, S^{\epsilon}_\mu (\xi)
   V^{\epsilon}_j(\mu|z) \ \ .
\end{align}
The definition of $V^\text{B}$ is modeled after the construction \eqref{phiBmuxi}
in the particle theory. From the discussion above we infer that
\begin{align}
 &\tilde C^\Hrep_b(j_1,j_2,j_3) \, D_{123}[j_i,\mu_i]  =  \nonumber \\
  & = 
 \ \lim_{z_{\infty} \to \infty} 
  \int \prod_{i=1}^3 d \bar \xi_i d \xi_i \ 
|z_{\infty}|^{4 \Delta^\Hrep_{j_1}} \langle V^{\text{B}}_{j_1}
  (\mu_1,\xi_1|z_\infty)
 V^{\text{B}}_{j_2} (\mu_2,\xi_2|1) V^{\text{B}}_{j_3}
(\mu_3,\xi_3|0) \rangle \nonumber \\[2mm]
&  = \ \lim_{z_{\infty} \to \infty} 
 \frac{1}{2 \sqrt2}
 \sum_{\epsilon_i = \pm} ( - \epsilon_1 \epsilon_2 \epsilon_3)
 |z_{\infty}|^{4 \Delta^\Hrep_{j_1}}
\langle V^{\epsilon_1}_{j_1} (\mu_1|z_\infty)
 V^{\epsilon_2}_{j_2} (\mu_2|1) V^{\epsilon_3}_{j_3}
   (\mu_3|0) \rangle  \label{sum3pt}\ \ .
\end{align}
Now we need to insert our result \eqref{3PTmue} along with
formulas for ${\cal I}$ and ${\cal H}$. After that we can compute the sum
on the right hand side of the previous equation. It is easy
to see that all terms involving the Liouville structure
constant $\tilde C^L_R$ cancel from the resulting
expression. The terms proportional to $C^L_R$ are
determined with the help of the following auxiliary
formula
\begin{align}
 &   \left( G( bp_1,bp_2,bp_3;z)+ \sqrt{1 - z}
   \, G(bp_1,-bp_2,-bp_3;z)\right)
  \times \nonumber \\[2mm]  & \qquad \times
  \left(G(bp_1,bp_2,bp_3;\bar z) + \sqrt{1 - \bar z}
   \, G(bp_1,-bp_2,-bp_3;\bar z)\right)
  \ =  \\[2mm]  \nonumber &  \qquad \qquad \qquad = \
   |z|^{\frac{1}{2b^2} + i p_1}
   | 1 - z|^{\frac{1}{2b^2} + i p_2}
   {}_2 {\cal F}_1 (-\tfrac14 + \tfrac{i}{2} p ,
 \tfrac{1}{4} +\tfrac{i}{2} p_{12}, \tfrac12 + i p_1 ;z ) ~,
\end{align}
where $p_{12} = p_1 + p_2 - p_3$ etc.\ and $p = p_1 + p_2 + p_3$.
The parameter $z$ takes the same value $z = 1+ \mu_2/\mu_3$ as
before. In the derivation of the formula we have used the
well known identities
\begin{align}
\tfrac12 \sin \kappa \cos \kappa
F(\tfrac54, \tfrac34 , \tfrac32 , \sin ^2 \kappa )=  \sin \tfrac12 \kappa  ~,
\qquad
 F(\tfrac14, - \tfrac14, \tfrac12, \sin ^2 \kappa ) = \cos \tfrac12 \kappa ~,
\end{align}
and
\begin{align}
 (c - b - 1) F(a,b,c;z) + b(1-z)F(a,b+1,c;z)
  \ = \ (c-1) F(a - 1,b,c-1;z) ~.
\end{align}
The resulting expression for the sum in eq.\ \eqref{sum3pt} is
of the form $\tilde C^\Hrep_b D_{123}$ with $D_{123}$ given by
formula \eqref{CGC3} if the structure function $\tilde C^\Hrep
(j_1,j_2,j_3)$ is introduced as
\begin{align} \label{tC} \nonumber
 \tilde C_b^\Hrep (j_1,j_2,j_3)  &= \ \frac{1}{2 \pi}\,
 C^L_R (\alpha_1 - \tfrac{1}{2b},\alpha_2,\alpha_3)
 \,  \frac{
  \gamma(\tfrac14 - \tfrac{i}{2} p) \gamma ( \tfrac12 + ip_1  )}
       { \gamma(\tfrac14 + \tfrac{i}{2} p_{31})
        \gamma(\tfrac14 + \tfrac{i}{2} p_{12})} \\[2mm]
   & =\ \frac{1}{2 \pi b}\,  
    \left( 
    \frac{2kb^{2 +b^2}}{i \gamma (\tfrac{b^2 + 1}{2})} \right)^{2j+5}
   \frac{\Upsilon ' _{\text{NS}}(0) }
  {\Upsilon_{\text{NS}}( 2 b(j + \frac52 ) + \frac{1}{b})}
  \times  \\[2mm]  & \times
 \frac{\Upsilon_{\text{R}}(4 b (j_1 + 1 ) + \frac{1}{b})
     \Upsilon_{\text{R} } (4 b (j_2 + 1 ) + \frac{1}{b} )
   \Upsilon_{\text{R}} (4 b (j_3 + 1 ) + \frac{1}{b}) }
     {
     \Upsilon_{\text{R} } (2 b(j_{12} + 1 ) + \frac{1}{b})
     \Upsilon_{\text{R}} (2 b(j_{23} + 1 ) + \frac{1}{b})
     \Upsilon_{\text{R}} (2 b(j_{31} + 1 ) + \frac{1}{b}) } ~.
     \nonumber
 \end{align}
A very similar analysis furnishes an expression for the structure
constant $C^\Hrep_b(j_1,j_2,j_3)$. Another glance onto eq.\ \eqref{3ptmuxi}
shows that $C^\Hrep_b$ may be determined e.g.\ from the terms proportional
to $\xi_1 \bar \xi_1 \xi_2 \bar \xi_2$ in the correlators of
$V_j^{\text{B}}(\mu,\xi|z)$,
\begin{align}
 & C^\Hrep_b(j_1,j_2,j_3) \, D_{12}[j_i,\mu_i] \nonumber \\
 & = \ \lim_{z_{\infty} \to \infty} 
  \int 
 \prod_{i=1}^3 d \bar \xi_i d  \xi_i \ \xi_3 \bar \xi_3
  |z_{\infty}|^{4 \Delta^\Hrep_{j_1}} \langle V^{\text{B}}_{j_1}
  (\mu_1,\xi_1|z_\infty)
 V^{\text{B}}_{j_2} (\mu_2,\xi_2|1) V^{\text{B}}_{j_3}
(\mu_3,\xi_3|0) \rangle \nonumber \\[2mm]
&  = \ \lim_{z_{\infty} \to \infty} 
 \frac{1}{2 \sqrt2}
 \sum_{\epsilon_i = \pm} \frac{\epsilon_1 \epsilon_2}{|\mu_3|}
 |z_{\infty}|^{4 \Delta^\Hrep_{j_1}}
\langle V^{\epsilon_1}_{j_1} (\mu_1|z_\infty)
 V^{\epsilon_2}_{j_2} (\mu_2|1) V^{\epsilon_3}_{j_3}
   (\mu_3|0) \rangle  \label{sum3pt2}\ \ .
\end{align}
The sum on the right
hand side can be computed in the precisely the same way as
before, and the result is given by replacing $p_3$ with
$- p_3$ and $C^L_R$ with $\tilde C^L_R$.
Thus we conclude
\begin{align}\label{C} \nonumber
 C_b^\Hrep (j_1,j_2,j_3)  &= \ \frac{1}{2 \pi}
 \, \tilde C^L_R (\alpha_1 - {\textstyle \frac{1}{2b}},
  \alpha_2,\alpha_3) \, \frac{
  \gamma(\tfrac34 - \tfrac{i}{2} p) \gamma ( \tfrac12 + ip_1  )}
       { \gamma(\tfrac34 + \tfrac{i}{2} p_{31})
        \gamma(\tfrac34 + \tfrac{i}{2} p_{12})} \\[2mm]
   & =\  \frac{1}{2 \pi}\,  \left( 
    \frac{2kb^{2 +b^2}}{i \gamma (\tfrac{b^2 + 1}{2})} \right)^{2j+5}
   \frac{\Upsilon ' _{\text{NS}}(0) }
  {\Upsilon_{\text{R}}( 2 b(j + \frac52 ) + \frac{1}{b})}
  \times  \\[2mm]  & \times
 \frac{\Upsilon_{\text{R}}(4 b (j_1 + 1 ) + \frac{1}{b})
     \Upsilon_{\text{R} } (4 b (j_2 + 1 ) + \frac{1}{b} )
   \Upsilon_{\text{R}} (4 b (j_3 + 1 ) + \frac{1}{b}) }
     {
     \Upsilon_{\text{NS} } (2 b(j_{12} + 1 ) + \frac{1}{b})
     \Upsilon_{\text{NS}} (2 b(j_{23} + 1 ) + \frac{1}{b})
     \Upsilon_{\text{NS}} (2 b(j_{31} + 1 ) + \frac{1}{b}) } ~.
     \nonumber
 \end{align}


\section{Conclusion}

In this note we have solved the very simplest example of a 
WZNW model on a type II supergroup, namely on the supergroup 
\OSP. Our discussion here was restricted to the NSNS sector 
of the theory but the analysis can easily be extended to the 
RR sector. The associated structure constants then involve 
the 2- and 3-point couplings in the NSNS sector of ${\cal N} 
=1$ Liouville theory. A more interesting problem would be to 
include boundary conditions into the analysis. According to 
\cite{Alekseev:1998mc} (see also \cite{Creutzig:2007jy} for 
a generalization to supergroups), maximally symmetric branes 
in the \OSP\ WZNW model correspond to (twisted) super-conjugacy 
classes. Under the \OSP\ WZNW super-Liouville correspondence, 
branes in the \OSP\ model should map to branes in ${\cal N}=1$ 
Liouville theory. The latter have been studied by several 
authors, see in particular \cite{FH,Ahn:2002ev}. In addition, 
it should also be possible to find a precise relation between 
correlation functions on the half-plane. In the case of the 
ordinary $H^+_3$-Liouville correspondence, such relations 
were found in \cite{HR} and rederived by means of the path 
integral approach in \cite{Fateev:2007wk}.%
\smallskip

To the best of our knowledge, the \OSP\ WZNW model had not 
been solved previously, though it is certainly possible to 
find its 2- and 3-point couplings more directly, i.e.\ without 
the relation to supersymmetric Liouville theory,  through the 
evaluation of factorization constraints. Such an approach has 
been successfully applied to the $H^+_3$ model in 
\cite{Teschner:1997ft}. It would be interesting to generalize 
the analysis of factorization constraints to the \OSP\ WZNW 
model.   
\smallskip 

The proposal of Ribault and Teschner for the concrete
relation between local correlators in $H^+_3$ model and 
Liouville field theory emerged partly from a careful 
comparison of differential equations on both sides of 
the correspondence. The correlators of any WZNW model 
obey the famous Knizhnik-Zamolodchikov equations. On the 
Liouville side, one higher order differential equation
arises from each degenerate field insertion. These two 
types of differential equations are mapped onto each 
other by the $H^+_3$-Liouville correspondence, see 
\cite{Ribault:2005wp,HS}. A similar analysis for the 
relation between the \OSP\ WZNW model and ${\cal N}=1$ 
Liouville theory has not been performed yet. 
\smallskip

There are various other extensions of our path integral 
approach that merit further study. 
Our basic strategy above was to apply the reduction ideas of 
\cite{HS} to the \tsl\ current algebra that resides within 
the \osp\ current algebra of the WZNW model. As we have 
explained in the introduction and illustrated in section 2.2, 
the same concepts apply to more general type II supergroups. 
It might be interesting to work this out in more detail, 
in particular for supergroups OSP($p|$N) with parameters 
N $\geq 3$. Another obvious extension would be to solve 
the OSP(2$|$2) WZNW model through its relation with ${\cal N}
=2$ Liouville field theory. While bulk 2-point functions of 
the latter model have been studied \cite{Ahn:2002sx} and a 
conjecture for bulk 3-point functions was formulated in 
\cite{Hosomichi:2004ph}, higher correlators are not yet available. In 
this context, it may also be worthwhile investigating the 
precise relation between the OSP(2$|$2) WZNW model discussed 
above and the SL(1$|$2) theory that has been solved in 
\cite{Saleur:2006tf,QuSc}. The OSP(2$|$2) WZNW model was
also investigated in the condensed matter literature, see
e.g.\ \cite{Maassarani:1996jn,Ludwig:2000em} and references 
therein.
\smallskip

In the case of WZNW models on type I supergroups it is 
possible to solve them in terms of a purely bosonic model
\cite{QuSc}. It seems likely that for type II supergroups
a further reduction is possible in which the remaining 
fermionic fields are also removed. Indeed, for ${\cal N}=1$ 
Liouville field theory the structure constants are very 
closely related to those of the purely bosonic Liouville 
model. It would be rewarding to find a formal path integral 
derivation of this relation  and to generalize it to higher 
supergroups.  
\smallskip 

Let us finally mention a rather different direction to which 
some of the above might apply. We have discussed in the 
introduction that correspondences of the proposed type 
elevate a usual Hamiltonian reduction to an equivalence
between local field theories of different  target space
dimension. But Hamiltonian reduction also links WZNW 
models for groups of higher rank to certain conformal 
Toda theories, see e.g.\ \cite{Forgacs:1989ac}. It is 
indeed likely that $N$-point functions 
of tachyon vertex operators in WZNW models can be more 
generally related to correlators in Toda theory. Unfortunately,  
no explicit formulas have been derived yet. The main 
technical obstacle arises from the non-abelian nature
of the maximal nilpotent subalgebra. In this sense, 
even the \osp\ case we have studied here could turn 
out to be a rather instructive example. The maximal 
nil-potent subalgebra of \osp, i.e.\ the algebra 
spanned by $F^+$ and $E^+$,  is non-abelian. Hence an 
equivalence 
between the \OSP\ WZNW model and bosonic Liouville
field theory (see previous paragraph) could be the 
first instance of a much more general class of 
dualities involving WZNW models on groups of rank 
$r > 1$ and conformal Toda theory. We plan to return 
to this subject in the near future.

\bigskip
\bigskip

\noindent {\bf Acknowledgments:} We are grateful to Thomas
Creutzig, Vladimir Mitev, Ioannis Papadimitriou, Sylvain Ribault
and J\"org Teschner for useful discussions and comments. The work
of YH was supported by an JSPS Postdoctoral Fellowship for
Research Abroad under contract number H18-143.


\appendix

\section{The Lie superalgebras \osp\ and sl(1$|$2)}

In this appendix we collect a few relevant details concerning the
two Lie superalgebras that feature in the main text, namely the
superalgebras \osp\ and sl(1$|$2).

\subsection{The Lie superalgebra \osp}
\label{App:osp}
The Lie superalgebra \osp\ possesses three bosonic generators
and two fermionic ones. We shall denote the former by $E^\pm,
H$ and use $F^\pm$ for fermionic generators. The relations
between these elements are given by
\begin{align}
 [ H , E^{\pm} ] &=\  \pm E^{\pm} ~,
&[ H , F^{\pm} ] &=\  \pm \frac{1}{2} F^{\pm} ~,
& [ E^+ , E^- ] &=\  2 H ~,
\\
 [ E^{\pm} , F^{\mp} ] &=\  - F^{\pm} ~,
 &\{ F^+ , F^- \} &=\  \frac{1}{2} H ~,
 &\{ F^{\pm} , F^{\pm} \} &=\  \pm \frac{1}{2} E^{\pm} ~. \nonumber
\end{align}
Note that $E^\pm$ and $H$ generate a \tsl\ subalgebra within
\osp. It is easy to verify that the following matrices provide
a 2$|$1-dimensional representation of \osp\ \cite{FSS},
\begin{align}
  H &=
 \begin{pmatrix}
  \frac{1}{2} & 0 & 0 \\
  0 & -\frac{1}{2} & 0 \\
  0 & 0 & 0
 \end{pmatrix} ~,
  &E^+ &=
 \begin{pmatrix}
  0 & 1 & 0 \\
  0 & 0 & 0 \\
  0 & 0 & 0
 \end{pmatrix} ~,
 &E^- &=
 \begin{pmatrix}
  0 & 0 & 0 \\
  1 & 0 & 0 \\
  0 & 0 & 0
 \end{pmatrix} ~, \\[3mm]
  F^+ &=
 \begin{pmatrix}
  0 & 0 & \frac{1}{2} \\
  0 & 0 & 0 \\
  0 & \frac{1}{2} & 0
 \end{pmatrix} ~,
 &F^- &=
 \begin{pmatrix}
  0 & 0 & 0 \\
  0 & 0 & - \frac{1}{2} \\
  \frac{1}{2} & 0 & 0
 \end{pmatrix} ~.\nonumber
\end{align}
The Lie superalgebra \osp\ possesses a non-degenerate invariant metric
$\langle X , Y \rangle = \text{str} (XY)$ which is defined for any pair
of elements $X,Y \in$ \osp\ and using the supertrace in the 2$|$1-dimensional
matrix representation,
\begin{align}
 \langle H , H \rangle &=\ \frac{1}{2} ~,
 & \langle E^+ , E^- \rangle &=\ \langle E^- , E^+ \rangle \ =\ 1 ~,
 & \langle F^+ , F^- \rangle &=\ - \langle F^- , F^+ \rangle \ = \
\frac{1}{2} ~.
\end{align}
The metric is needed e.g.\ to write down the action of a WZNW model
on the Lie supergroup \OSP.

\subsection{The Lie superalgebra osp(2$|$2)}
\label{App:sl12}

The Lie superalgebra possesses four bosonic generators $E^\pm, H$ and
$Z$ along with four fermionic ones. The latter are denotes by $F^\pm$
and $\bar F^\pm$. These eight generators obey the following set of
non-trivial (anti-)commutation relations \cite{FSS},
\begin{align}
 [ H , E^{\pm} ] &= \pm E^{\pm} ~,
&[ H , F^{\pm} ] &= \pm \frac{1}{2} F^{\pm} ~,
&[ H , \bar F^{\pm} ] &= \pm \frac{1}{2} \bar F^{\pm} ~,  \nonumber\\
[ Z , F^{\pm} ] &=  \frac{1}{2} F^{\pm} ~,
&[ Z , \bar F^{\pm} ] &= - \frac{1}{2} \bar F^{\pm} ~,
 &[ E^+ , E^- ] &= 2 H ~,\\[2mm]
[ E^{\pm} , F^{\mp} ] &= - F^{\pm} ~,
 &[ E^{\pm} , \bar  F^{\mp} ] &= \bar  F^{\pm} ~,
 &\{ F^{\pm} , \bar F^{\mp} \} &= Z \mp H ~,
 &\{ F^{\pm} , \bar F^{\pm} \} &= E^{\pm} ~.\nonumber
\end{align}
As in the case of \osp\ it is possible to find a matrix representation
of osp(2$|$2) that is built out of 2$|$1-dimensional supermatrices,
\begin{align}
  H &=
 \begin{pmatrix}
  \frac{1}{2} & 0 & 0 \\
  0 & -\frac{1}{2} & 0 \\
  0 & 0 & 0
 \end{pmatrix} ~,
  &Z &=
 \begin{pmatrix}
  \frac{1}{2} & 0 & 0 \\
  0 & \frac{1}{2} & 0 \\
  0 & 0 & 1
 \end{pmatrix} ~,
  &E^+ &=
 \begin{pmatrix}
  0 & 1 & 0 \\
  0 & 0 & 0 \\
  0 & 0 & 0
 \end{pmatrix} ~,
 &E^- &=
 \begin{pmatrix}
  0 & 0 & 0 \\
  1 & 0 & 0 \\
  0 & 0 & 0
 \end{pmatrix} ~, \\
  F^+ &=
 \begin{pmatrix}
  0 & 0 & 0 \\
  0 & 0 & 0 \\
  0 & 1 & 0
 \end{pmatrix} ~,
  &\bar F^+ &=
 \begin{pmatrix}
  0 & 0 & 1 \\
  0 & 0 & 0 \\
  0 & 0 & 0
 \end{pmatrix} ~,
 &F^- &=
 \begin{pmatrix}
  0 & 0 & 0 \\
  0 & 0 & 0 \\
  1 & 0 & 0
 \end{pmatrix} ~,
 &\bar F^- &=
 \begin{pmatrix}
  0 & 0 & 0 \\
  0 & 0 & 1 \\
  0 & 0 & 0
 \end{pmatrix} ~. \nonumber
\end{align}
Using the prescription $\langle X,Y\rangle = \text{str} (XY)$ it is
easy to find the following non-trivial bilinear form on osp(2$|$2),
\begin{align}
 &\langle  H , H  \rangle = \frac{1}{2} ~, \qquad
 \langle Z , Z \rangle = - \frac{1}{2} ~, \qquad
 \langle E^+ , E^- \rangle = \langle E^- ,  E^+ \rangle = 1 ~,
 \\[2mm]
&\langle F^+ , \bar F^- \rangle = \langle\bar F^+ , F^- \rangle =
1 ~, \qquad \langle \bar F^- , F^+ \rangle = \langle F^- , \bar
F^+ \rangle
 = - 1~. \nonumber
\end{align}
The form $\langle \cdot,\cdot\rangle$ defines a non-degenerate invariant
metric on the Lie superalgebra osp(2$|$2).

\section{Integral formulas}
\label{int}

In this appendix we list a few simple integral formulas that are
used in some of the derivations we sketched in the main part of
this note.
\smallskip

The first formula concerns the overlap of two Bessel functions
that is needed in section 3.1 on the minisuperspace theory.
Utilizing the formula
\begin{align}
 &\int _0^\infty dx x^{\alpha - 1} K_{\mu} (x) K_\nu (x)
  = \nonumber \\[2mm]
  &=\ \frac{2^{\alpha -3}}{\Gamma(\alpha)}
  \Gamma\left(\frac{\alpha +\mu + \nu}{2}\right)
  \Gamma\left(\frac{\alpha +\mu - \nu}{2}\right)
  \Gamma\left(\frac{\alpha -\mu + \nu}{2}\right)
  \Gamma\left(\frac{\alpha -\mu - \nu}{2}\right)  ~,
\end{align}
and
\begin{align}
 \Gamma (ix) \Gamma (1 - i x) &=\ \frac{\pi}{i \sinh \pi x} ~,
 &\Gamma (\tfrac12 + ix)  \Gamma (\tfrac12 - ix)
   &=\ \frac{\pi}{\cosh \pi x} ~,
\end{align}
we obtain
\begin{align}
 &\int_0^\infty d x \left( K_{\frac{1 + \epsilon}{2} + i p} ( x)
  \pm K_{\frac{1 + \epsilon}{2} - i p} (x)  \right)
   \left( K_{\frac{1 + \epsilon}{2} + i p '} ( x)
  \pm K_{\frac{1 + \epsilon}{2} - i p '} ( x)  \right)
   \nonumber \\[2mm]
 &=\ \frac{1}{4 i} \Biggl[ \left(
   \frac{\pi}{\sinh \pi (\frac{p + p'}{2} - i \epsilon )} -
    \frac{\pi}{\sinh \pi (\frac{p + p'}{2} + i \epsilon )}  \right)
   \frac{\pi}{\cosh \pi (\frac{p-p'}{2})}   \nonumber \\[2mm]
 & \quad \pm \left(
   \frac{\pi}{\sinh \pi (\frac{p - p'}{2} - i \epsilon )} -
    \frac{\pi}{\sinh \pi (\frac{p - p'}{2} + i \epsilon )}  \right)
     \frac{\pi}{\cosh \pi (\frac{p+p'}{2})}  \Biggr] ~.
\end{align}
If we take $\epsilon \to 0$, then the above quantity vanishes except for
$p = \pm p'$. Around these points we may use
\begin{align}
  \frac{\pi}{\sinh \pi (\frac{p + p'}{2} - i \epsilon )} -
    \frac{\pi}{\sinh \pi (\frac{p + p'}{2} + i \epsilon )}
    \sim \frac{2 i \epsilon}{(\frac{p + p'}{2})^2 + \epsilon ^2 }
    \to 4 \pi i \delta (p+p') ~.
\end{align}
These results are exploited in our computation \eqref{2ptmue} of the
particle 2-point correlator.
\smallskip

In passing from the 2-point function \eqref{2ptxxi} in the $(x,\xi)$
basis to the $(\mu,\epsilon)$ basis, we make use of the following
simple integrals,
\begin{align}
  & \frac{1}{2 \pi^2} \int \prod_{i=1,2}
  \left[ |\mu_i|^{2j_i + 2} d^2 x_i d \bar \xi_i d \xi_i 
   ( 1 - \epsilon_i |\mu_i| \xi_i \bar \xi_i  )
  e^{ \mu_i x_i - \bar \mu_i \bar x_i} \right]
  \delta^2 (x_1 - x_2) (\xi_1 - \xi_2 ) ( \bar \xi_1 - \bar \xi_2 )
 \ = \nonumber  \\[2mm]
 & \qquad =  \ - 4  |\mu_2 |^2 \epsilon_2
 \delta_{\epsilon_1,\epsilon_2} \delta ^2 (\mu_1 + \mu_2 ) 
\end{align}
with $j_1 + j_2 + 3/2 = 0$, and
\begin{align}
  & \frac{1}{2 \pi^2}
 \int\!\prod_{i=1,2} \left[ |\mu_i|^{2j_i + 2} d ^2 x_i d \bar \xi_i d \xi_i
  ( 1 - \epsilon_i |\mu_i | \xi_i \bar \xi_i  )
  e^{ \mu_i x_i - \bar \mu_i \bar x_i} \right]
  ( 1 +   \xi_1 \xi_2 \bar \xi_1 \bar \xi_2 \partial_{x_1} \partial_{\bar x_1})
  | x_1 - x_2 |^{4 j_1 + 2} = \nonumber  \\[2mm]
 & \qquad = \  4 | \mu_2 | ^2
  \delta_{\epsilon_1,\epsilon_2} \delta ^2 (\mu_1 + \mu_2 )
\pi \gamma ( 2 j_2 + 2) 
\end{align}
with $j_1 - j_2 = 0$.
Both formulas are straightforward to derive using only standard
properties of Grassmann integrals.

When computing three point functions, we use the function $D$
that is defined by the following integral formula
\begin{align}
&D \left[ \begin{smallmatrix}
j_1 &j_2 & j_3 \\
\mu_1 & \mu_2 & \mu_3
\end{smallmatrix}\right]
   = \frac{1}{\pi ^3}
 \int \prod_{i=1}^3 \left[ |\mu_i|^{2j_i + 2}
  d^2 x_i  e^{\mu_i x_i - \bar \mu_i \bar x_i}
 \right]
   |x_{12}|^{2 j_{12}} |x_{23}|^{2 j_{23}} |x_{31}|^{2 j_{31}}  ~,
   \label{CGC2}
\end{align}
where $j_{12} = j_1 + j_2 - j_3$ and so on. 
The integrations may be performed explicitly and they lead to a
rather bulky expression in terms of hypergeometric functions,
\begin{align}
D \left[ \begin{smallmatrix}
j_1 &j_2 & j_3 \\
\mu_1 & \mu_2 & \mu_3
\end{smallmatrix}\right]
   \label{CGC}
  & = \ \pi \delta^{(2)} (\mu_1+\mu_2+\mu_3)\, | \mu_3|^{ - 2 j_1 - 2 j_2 - 2}
   |\mu_1|^{2j_1 + 2} |\mu_2|^{2j_2+2}
   \times  \\[2mm] & \times
      \left[ \frac{\gamma (j_{31} + 1) \gamma (j_{12} + 1)}
      {\gamma(- j - 1 ) \gamma (2 j_1 + 2)  }
    {}_2 {\cal F }_1 (j + 2 , j_{12} + 1 , 2 j_1 + 2 ;
     1 + \tfrac{\mu_2}{\mu_3}  ) \right. \nonumber \\[2mm]
    &  \qquad \left.
    + | 1 + \tfrac{\mu_2}{\mu_3} |^{-2(2 j_1 + 1)}
     \frac{\gamma(j_{23} + 1 )}{\gamma(-2 j_1)}
    {}_2 {\cal F }_1 ( - j_{31} , j_{23} + 1 , - 2 j_1;
     1 + \tfrac{\mu_2}{\mu_3}  )  \right] ~.
    \nonumber
\end{align}
Here we have used $j = j_1 + j_2 + j_3$ and
\begin{align}
 {}_2 {\cal F }_1 ( a ,b , c ; z ) = F(a,b,c;z) F(a,b,c;\bar z) ~.
\end{align}

\section{${\cal N}=1$ super Liouville theory}
\label{Liouville}

In order to carry out the computations of section 4, we need 
rather extensive information on correlation functions in ${\cal N}=1$ 
Liouville field theory. For the convenience of the reader we 
collect all relevant formulas in this appendix. Most of the 
results are taken from \cite{FH}. 
\smallskip 

In our conventions, the action of ${\cal N}=1$ super Liouville field 
theory takes the form 
\begin{align}
   S^{L}
   \ = \ \frac{1}{4\pi} \int d^2 z \left[  \partial \varphi \bar 
   \partial \varphi
   + \frac{Q}{4} \sqrt g R \varphi
  + \psi \bar \partial \psi + \bar \psi \partial \bar \psi \right] +
  i \mu_L b^2 \int d^2 z \psi \bar \psi e^{ b \varphi}  ~,
\end{align}
where $Q = b + 1/b$. For the relation with the \OSP\  WZNW model
at level $k$ we set $b= 1/\sqrt{2 k - 3}$ and fix the bulk 
cosmological constant to be $\mu_L = i/(2\pi k b^2)$. 
\smallskip 

As all fermionic models, ${\cal N}=1$ Liouville theory possesses two 
sectors. Depending on the boundary conditions on fermions, these
are denoted by NSNS (Neveu-Schwarz) and RR (Ramond) sectors. Primary 
fields in the NSNS sector can be thought of as exponentials 
$V_{\alpha} = e^{\alpha \varphi}$ in the bosonic field $\varphi$. 
Their conformal weight is given by $\Delta_\alpha ^L = \alpha (Q - 
\alpha)/2$. The 2-point function of these NSNS primary fields 
takes the form  
\begin{align}\label{L2ptNS} 
 \langle V_{\alpha_1} (z_1) V_{\alpha_2} (z_2 ) \rangle 
\  =\  |z_{12}|^{-4 \Delta^L_{\alpha_2}}
  2 \pi \left[ \delta (\alpha_1 + \alpha_2 - Q)
+ \delta (\alpha_1 - \alpha_2 ) D^L_{NS}(\alpha_2 ) \right] ~,
\end{align}
with
\begin{align}
 D^L_{NS} (\alpha) \ = \ - \left(\mu_L \pi \gamma 
    (\tfrac{bQ}{2}) \right)^{\frac{Q-2\alpha}{b}}
  \frac{\Gamma ( b (\alpha-\frac{Q}{2}) )
        \Gamma ( \frac{1}{b} (\alpha-\frac{Q}{2}) ) }
       {\Gamma (- b (\alpha-\frac{Q}{2}) )
        \Gamma ( - \frac{1}{b} (\alpha-\frac{Q}{2}) ) } ~.
\end{align}
Here and throughout the main text we use $\gamma(x) = \Gamma (x)/
\Gamma(1-x)$. Whereas the first term in eq.\ \eqref{L2pt} is fixed 
by normalization, the second term involving  $D^L_{NS}$
contains dynamical information on the phase shift of tachyonic 
modes upon reflection off the Liouville wall. 
\smallskip 

The vertex operators that appear in our relation with the \OSP\ 
WZNW model, and in particular in eq.\ \eqref{sRTeven}, are all 
in the RR sector. Before we can spell our properties of RR-fields
we want to recall a few basic facts on spin fields which apply to 
${\cal N}=1$ Liouville theory and free fermions alike. Chiral spin fields 
$\sigma^{\pm}$ and  $\bar \sigma^{\pm}$ may be characterized by 
their operator product with the fermions,
\begin{align}
 \psi (z) \sigma^{\pm} (0) &\sim \frac{\sigma^{\mp}(0)}{\sqrt2 z^{\frac{1}{2}}} ~,
& \bar \sigma^{\pm} (\bar z) \bar \psi (0)
  &\sim \frac{i \bar \sigma^{\mp}(0)}{\sqrt2 \bar z^{\frac{1}{2}}} ~.
\end{align}
As usual, we combine left- and right-moving spin fields into the 
non-chiral products $\sigma^{\epsilon \bar \epsilon} = \sigma^\epsilon 
\bar \sigma^{\bar \epsilon}$. Their operator products are known to 
be given by 
\begin{align}
 \sigma^{\pm \pm} (z) \sigma^{\pm \pm} (0)&\sim\ \frac{1}{|z|^{\frac14}} ~,
& \sigma^{\pm \mp} (z) \sigma^{\pm \mp} (0)&\sim\ \frac{i}{|z|^{\frac14}} ~, 
  \\[2mm] 
 \sigma^{\pm \pm} (z) \sigma^{\mp \mp}(0)
    &\sim\ - \frac{i}{2} \psi \bar \psi(0) |z|^{\frac34}~,
& \sigma^{\pm \mp} (z) \sigma^{\mp \pm}(0)
    &\sim\ - \frac{1}{2} \psi \bar \psi(0) |z|^{\frac34}~.
\end{align}
Only two special linear combinations of the spin fields play an 
important role for the theory. These are introduced as follows 
\begin{align}
 \Sigma^+ &=\ \frac{1}{\sqrt2} (\sigma^{++} - \sigma^{--}) ~,
 &\Sigma^- &=\ \frac{e^{-\pi i /4}}{\sqrt2}
  (\sigma^{-+} - \sigma^{+-} ) ~. 
\end{align}
In the case of the Ising model (free fermions), $\Sigma^+ (=\sigma)$ 
is known as the order field while $\Sigma^- (=\mu)$ is referred to 
as the disorder field. From the operator products of spin fields we 
conclude easily, 
\begin{align}
 \Sigma^+ (z) \Sigma^+ (0) &\sim\ \frac{1}{|z|^{\frac14}}
  + \frac{i}{2} \psi \bar \psi |z|^{\frac34} ~,
 &\Sigma^- (z) \Sigma^- (0) &\sim\ \frac{1}{|z|^{\frac14}}
  - \frac{i}{2} \psi \bar \psi |z|^{\frac34} ~.
\end{align}
${\cal N}=1$ Liouville field theory contains a family of spin fields 
which is parametrized by their `momentum' $\alpha$ in the $\varphi$ 
direction. We can think of these primary fields in the RR sector as
products of a spin field and an exponential,  
\begin{align}
 \Theta^{\epsilon \bar \epsilon}_{\alpha} &=\ 
 \sigma^{\epsilon \bar \epsilon}  e^{ \alpha \varphi} ~,
 &\Sigma^{\pm}_{\alpha} &=\ \Sigma^{\pm} e^{ \alpha \varphi} ~.
 \label{spin}
\end{align}
The 2-point functions of the vertex operators $\Sigma^{\epsilon}_{\alpha}$ 
possess the following form 
\begin{align}
&\langle \Sigma^{\pm}_{\alpha_1} (z_1) \Sigma^{\pm}_{\alpha_2} (z_2) \rangle
\ = \ |z_{12}|^{- 4 \Delta ^L _{\alpha_2} - \frac14} 2 \pi
 \left[\delta (\alpha_1 + \alpha_2 - Q) \mp \delta (\alpha_1 - \alpha_2 )
  D^L_R  (\alpha_2) \right] ~
 \label{L2pt}
\end{align}
with a reflection coefficient given by 
\begin{align}
 D^L_R (\alpha )\  = \  
 \left(\mu_L \pi \gamma (\tfrac{bQ}{2}) \right)^{\frac{Q-2\alpha}{b}}
  \frac{\Gamma (\frac12 + b (\alpha-\frac{Q}{2}) )
        \Gamma (\frac12 + \frac{1}{b} (\alpha-\frac{Q}{2}) ) }
       {\Gamma (\frac12 - b (\alpha-\frac{Q}{2}) )
        \Gamma (\frac12 - \frac{1}{b} (\alpha-\frac{Q}{2}) ) } ~.
        \label{refR}
\end{align}
In order to compute 3-point functions of \OSP\ model,
we need 4-point functions of RR-fields involving a single 
degenerate field $\Theta^{\pm}_{-1/2b}$ in ${\cal N}=1$ super 
Liouville theory. The states $|-1/2b\rangle_\pm = |1,2\rangle_\pm$  
that correspond to the degenerate field are characterized by the 
relations, 
\begin{align}
 G_0 |1,2\rangle_{\pm} &=\ 
 \frac{i}{\sqrt2} \left[\frac{1}{b} + \frac{b}{2} \right]|1,2\rangle_{\mp} ~,
& G_{-1}  |1,2\rangle_{\pm} + i \sqrt2 b L_{-1} |1,2\rangle_{\mp} &=\ 0 ~.
\label{BPZeq}
\end{align}
It follows from standard arguments that operator products involving 
at least one such degenerate field contain two terms only, 
\begin{align}\nonumber
 - 2 \Theta^{\pm \mp}_{-1/2b} (z_1) \Theta^{\mp \pm}_\alpha (z_2)
   &\sim |z_{12}|^{\frac{ \alpha}{b} + \frac{3}{4}}
   \psi \bar \psi V_{\alpha - 1/2b}  (z_2)
  + C^L_{R,-} (\alpha) |z_{12}|^{\frac{1}{b}(Q-\alpha) -\frac14 }
    V_{\alpha+1/2b}  (z_2)~,
\end{align}
where
\begin{align}
 C^L_{R,-} (\alpha) \ = \ 2 i   D^L_{R} (\alpha) D^L_{NS} (Q - \alpha - \tfrac{1}{2b})
 \  =\  2 i b^{-2}
 \left(\mu_L \pi \gamma ( \tfrac{b Q}{2} ) \right)^{\frac{1}{b^2}}
   \gamma (\tfrac12 - \tfrac{ \alpha}{b})
   \gamma (\tfrac{ \alpha}{b} - \tfrac{1}{2 b^2} ) ~.
\nonumber \end{align}
The other operator product expansions we will need below can be obtained 
from the one we have provided by super conformal transformation along 
with the relations
\begin{align}
 \Theta^{\epsilon_1 , \bar \epsilon_1}_{\alpha_1} (z_1)
 \Theta^{\epsilon_2 , \bar \epsilon_2}_{\alpha_2} (z_2)
  \sim - i \bar \epsilon_1 \epsilon_2
\Theta^{-\epsilon_1 , \bar \epsilon_1}_{\alpha_1} (z_1)
 \Theta^{-\epsilon_2 , \bar \epsilon_2}_{\alpha_2} (z_2)
  \sim - i \bar \epsilon_1 \epsilon_2
\Theta^{\epsilon_1 , -\bar \epsilon_1}_{\alpha_1} (z_1)
 \Theta^{\epsilon_2 , -\bar \epsilon_2}_{\alpha_2} (z_2) ~.
 \label{signrel}
\end{align}

Before we spell out a formula for the relevant 4-point functions, 
let us provide explicit expressions for the 3-point functions 
involving two RR fields. These were determined in 
\cite{RS,Poghosian,FH} and we shall simply quote their results 
along with all the necessary notations, 
\begin{align}
 \langle V_{\alpha_1} (z_1) \Theta_{\alpha_2}^{\pm \pm} (z_2)
 \Theta_{\alpha_3}^{\mp \mp} (z_3)
 \rangle
 \ = \ |z_{12}|^{-2\Delta^{L}_{12}}
   |z_{23}|^{-2\Delta^{L}_{23} - \frac14}
   |z_{31}|^{-2\Delta^{L}_{31}}
    C^L_R (\alpha_1 ; \alpha_2 , \alpha_3 ) ~, \\[2mm] 
 \langle V_{\alpha_1} (z_1) \Theta_{\alpha_2}^{\pm \pm} (z_2)
 \Theta_{\alpha_3}^{\pm \pm} (z_3)
 \rangle
 \ = \ |z_{12}|^{-2\Delta^{L}_{12}}
   |z_{23}|^{-2\Delta^{L}_{23}- \frac14}
   |z_{31}|^{-2\Delta^{L}_{31}}
    \tilde C^L_R (\alpha_1 ; \alpha_2 , \alpha_3 ) ~, 
\end{align}
where $\Delta^L_{12} = \Delta^L_{\alpha_1} + \Delta^L_{\alpha_2} 
- \Delta^L_{\alpha_3}$
etc. Once more, other 3-point functions may be obtained with the 
help of the relations \eqref{signrel}. The structure constants 
$C^L_R$ and $\tilde C^L_R$ are constructed from 
a special functions $\Upsilon$ as follows, 
\begin{align}\nonumber 
 &C^L_R (\alpha_1 ; \alpha_2 , \alpha_3 )
 \, = \, \left( \mu_L \pi \gamma (\tfrac{bQ}{2} )  b^{1-b^2}
    \right)^{\frac{Q- \alpha}{b}}
    \frac{\Upsilon ' _{\text{NS}}(0) \Upsilon_{\text{NS}}(2 \alpha_1)
     \Upsilon_{\text{R} } (2 \alpha_2 ) \Upsilon_{\text{R}} (2 \alpha_3) }
     { \Upsilon_{\text{R}}( \alpha - Q)
     \Upsilon_{\text{R} } (\alpha_{23})
     \Upsilon_{\text{NS}} (\alpha_{12})
     \Upsilon_{\text{NS}} (\alpha_{31}) } ~,\\[2mm] 
 &\tilde C^L_R (\alpha_1 ; \alpha_2 , \alpha_3 )
  \, =\  \left( \mu_L \pi \gamma ( \tfrac{bQ}{2} )  b^{1-b^2}
    \right)^{\frac{Q- \alpha}{b}}
    \frac{\Upsilon ' _{\text{NS}}(0) \Upsilon_{\text{NS}}(2 \alpha_1)
     \Upsilon_{\text{R} } (2 \alpha_2 ) \Upsilon_{\text{R}} (2 \alpha_3) }
     { \Upsilon_{\text{NS}}( \alpha - Q)
     \Upsilon_{\text{NS} } (\alpha_{23})
     \Upsilon_{\text{R}} (\alpha_{12})
     \Upsilon_{\text{R}} (\alpha_{31}) } ~,
\nonumber 
\end{align}
where $\alpha_{12} = \alpha_1 + \alpha_2 - \alpha_3$ etc., 
$\alpha = \alpha_1 + \alpha_2 + \alpha_3$, and 
\begin{align}
 \Upsilon_{\text{NS}} (x) &= \Upsilon ( \tfrac{x}{2})
   \Upsilon (\tfrac{x+Q}{2}) ~,
 &\Upsilon_{\text{R}} (x) &= \Upsilon (\tfrac{x+b}{2})
   \Upsilon (\tfrac{x+b^{-1}}{2}) ~.
\end{align}
The $\Upsilon$ function itself is closely related to Barnes 
double Gamma function. Instead of describing the precise connection, 
we simply display an integral representation
\begin{align}
 \ln \Upsilon (x) \ = \ \int_0^{\infty} \frac{dt}{t}
 \left[ e^{-2t} \left( \frac{Q}{2} - x\right)^2 -
   \frac{\sinh ^2 (\frac{Q}{2} - x) t}{\sinh bt \sinh \frac{t}{b} } \right] ~.
\end{align}
Note that the functions $\Upsilon_{\text{NS}}$ and $\Upsilon_{\text{R}}$ 
possess the following behavior under shifts of their argument, 
\begin{align}
 \Upsilon_{\text{NS}} (x+b) &= b^{-b x} \gamma ( \tfrac12 + \tfrac{bx}{2})
 \Upsilon_{\text{R}}(x) ~,
 &\Upsilon_{\text{R}} (x+b) &= b^{1-b x} \gamma (\tfrac{bx}{2})
 \Upsilon_{\text{NS}} (x)~, \\
  \Upsilon_{\text{NS}} (x+ \tfrac{1}{b})
  &= b^{\frac{x}{b}} \gamma (\tfrac12 + \tfrac{x}{2b})
 \Upsilon_{\text{R}} (x)~,
 &\Upsilon_{\text{R}} (x+ \tfrac{1}{b}) &= b^{-1+ \frac{x}{b} }
 \gamma (\tfrac{x}{2b})
 \Upsilon_{\text{NS}} (x) ~.
 \label{uprel}
\end{align}

Let us finally turn to a discussion of the 4-point functions 
involving one degenerate field along with three primary fields
from the RR sector. This quantity was computed in \cite{FH} and
it takes the form 
\begin{align}
 &\langle \Sigma^{\epsilon_0}_{-1/2b} (z_0)
  \Sigma^{\epsilon_1}_{\alpha_1} (z_1) \Sigma^{\epsilon_2}_{\alpha_2} (z_2)
 \Sigma^{\epsilon_3}_{\alpha_3} (z_3) \rangle
  = |z_{03}|^{-4 \Delta^L_{-1/2b} -\frac{1}{4}}
    \times \nonumber \\[2mm]  & \qquad \times
    |z_{12} |^{- 2 \Delta^{L}_{12} - \frac{1}{4} - 2 \Delta^L_{-1/2b} }
    |z_{23} |^{- 2 \Delta^{L}_{23} + 2 \Delta^L_{-1/2b} }
    |z_{31} |^{- 2 \Delta^{L}_{31}  + 2 \Delta^L_{-1/2b}}
     {\cal H}^{\epsilon_0 \epsilon_1 \epsilon_2 \epsilon_3}(z) ~,
     \label{L4pt}
\end{align}
where  ${\cal H}^{\epsilon_0 \epsilon_1 \epsilon_2 \epsilon_3}(z)$
is a function of the cross ratio $z = (z_{01} z_{23})/(z_{03} z_{21} )$.
We need some preparation before we can specify the functions ${\cal H}$. They 
are built from yet another set of auxiliary functions which depend on 
$\alpha_i = Q/2+ip_i$ according to  
\begin{align}
 &{\cal G}_0 (p_1,p_2,p_3;z) \ =\ 
  - \tfrac12 (z(1-z))^{\frac58}
  F(\tfrac{5}{4},\tfrac{3}{4},\tfrac{3}{2}, z)
  (G (p_1 , p_2 , p_3 ; z ) - G(p_1 , - p_2 , - p_3 ; z ) ) ~,
   \nonumber \\[2mm] 
 &{\cal G}_1 (p_1,p_2,p_3;z)\  =\ 
   (z(1-z))^{\frac18} 
   F(\tfrac{1}{4},-\tfrac{1}{4},\tfrac{1}{2}, z)
  (G (p_1 , p_2 , p_3 ; z ) + G(p_1 , - p_2 , - p_3 ; z ) ) ~,
\nonumber
\end{align}
where
\begin{align}
 G(p_1, p_2 , p_3 ; z) &=\ 
  z^{\frac{1}{4 b^2} + \frac{i p_1}{2 b}}
    (1 - z)^{\frac{1}{4 b^2} + \tfrac{i p_2}{2 b}}
  \left( \frac{\frac{1}{4} + \tfrac{i}{2 b} p_{31}}
  {\frac{1}{2}+ \tfrac{i}{b} p_1}\right)
   F(\tfrac{3}{4} + \tfrac{i}{2b} p,
     \tfrac{1}{4} + \tfrac{i}{2b} p_{12},
     \tfrac{3}{2} + \tfrac{i}{b} p_1 ; z ) ~ \nonumber
\end{align}
and $p_{12} = p_1 + p_2 - p_3$ etc., $p = p_1 + p_2 + p_3$.
One may show that ${\cal H}$ and ${\cal G}_a$ 
obey the same linear differential equations. Hence, we will be 
able to construct ${\cal H}$ from ${\cal G}_0, {\cal G}_1$ and 
\begin{align}
 {\cal G}_2 (p_1,p_2,p_3;z)& = \  {\cal G}_0 (- p_1,p_2,- p_3;z) ~,
& {\cal G}_3 (p_1,p_2,p_3;z) &= \  {\cal G}_1 (- p_1,p_2,- p_3;z) ~.
\end{align}
Combinations of these four functions ${\cal G}_a$ with trivial
monodromies around $z=0$ and $z=1$ are given by (see also \cite{FH}) 
\begin{align}
 H^{\pm}_1(p_1,p_2,p_3;z)
  & =\  (\pm {\cal G}_0 \bar {\cal G}_0 + {\cal G}_1 \bar {\cal G}_1 )
   + \gamma (\tfrac12 + \tfrac{i}{b} p_1)^2 \gamma (\tfrac14 + \tfrac{i}{2b}p_{23})
   \times  \\[2mm]  & \times
   \gamma (\tfrac14 - \tfrac{i}{2b}p)
   \gamma (\tfrac34 - \tfrac{i}{2b}p_{31})
   \gamma (\tfrac34 - \tfrac{i}{2b}p_{12})
    ( {\cal G}_2 \bar {\cal G}_2 \pm {\cal G}_3 \bar {\cal G}_3 ) ~,
    \nonumber \\[3mm] 
 H^{\pm}_2 (p_1,p_2,p_3;z)
  & =\  ( {\cal G}_0 \bar {\cal G}_1 \pm {\cal G}_1 \bar {\cal G}_0 )
   + \gamma (\tfrac12 + \tfrac{i}{b} p_1)^2 \gamma (\tfrac14 + \tfrac{i}{2b}p_{23})
   \times  \\[2mm] & \times
   \gamma (\tfrac14 - \tfrac{i}{2b}p)
   \gamma (\tfrac34 - \tfrac{i}{2b}p_{13})
   \gamma (\tfrac34 - \tfrac{i}{2b}p_{12})
    ( \pm {\cal G}_2 \bar {\cal G}_3 + {\cal G}_3 \bar {\cal G}_2 ) ~,
    \nonumber
\end{align}
where $\bar {\cal G}_i (p_1,p_2,p_3;z) = {\cal G}_i (p_1,p_2,p_3;\bar z)$.
Both $H^+_1$ and $H^-_2$ have previously appeared in \cite{FH} where they 
were also shown to be invariant under the crossing symmetry transformation 
$z \mapsto 1-z$ (note that $H^-_2$ flips its sign). Under the action of the
same crossing symmetry transformation, our functions $H^-_1$ and $H^+_2$
are mapped onto each other. We finally combine the functions $H_1$ and 
$H_2$ into ${\cal H}$, in a way that is determined by the 
desired operator product expansions, \def\talpha{{\tilde \alpha}}
\begin{align}
 & 2 {\cal H}^{\pm\pm\pm\pm} (z)
 \, = \, \mp C^L_R (\talpha_1,\alpha_2,\alpha_3) H^+_1(p_1,p_2,p_3;z)
  + \tilde  C^L_R (\talpha_1,\alpha_2,\alpha_3)
   H^+_1(p_1,p_2,- p_3;z) ~, \nonumber \\[2mm] 
 & 2 {\cal H}^{\pm\pm\mp\mp} (z)
  \, =\, \pm C^L_R (\talpha_1,\alpha_2,\alpha_3) H^-_1(p_1,p_2,p_3;z)
  + \tilde  C^L_R (\talpha_1,\alpha_2,\alpha_3)
   H^-_1(p_1,p_2,- p_3;z) ~, \label{feeee}
 \\[-2mm] 
 & 2 {\cal H}^{\pm\mp\pm\mp} (z)
  \, = \, -  C^L_R (\talpha_1,\alpha_2,\alpha_3)
    H^+_2 (p_1,p_2,p_3;z)
  \pm  \tilde  C^L_R (\talpha_1,\alpha_2,\alpha_3)
   H^+_2 (p_1,p_2,- p_3;z) ~,\nonumber \\[2mm] 
 & 2 {\cal H}^{\pm\mp\mp\pm} (z)
  \, =\, \mp i C^L_R (\talpha_1,\alpha_2,\alpha_3)
  H^-_2 (p_1,p_2,p_3;z)
  + i \tilde  C^L_R (\talpha_1,\alpha_2,\alpha_3)
   H^-_2 (p_1,p_2,- p_3;z) ~,
   \nonumber 
\end{align}
where the first argument of $C^L_R$ and $\tilde C^L_R$ is shifted by $1/2b$, 
i.e.\ we have set $\talpha = \alpha-1/2b$. This concludes our 
description of the 4-point functions.



\begin{thebibliography}{99}
\addtolength{\itemsep}{-3mm}

\bibitem{Metsaev:1998it}
  R.~R.~Metsaev and A.~A.~Tseytlin,
  {\it Type IIB superstring action in $AdS_5 \times S^5$ background},
  \npb{533}{1998}{109} [\hepth{9805028}].

\bibitem{Efetov}
  K.~B.~Efetov,
  {\it Supersymmetry and theory of disordered metals},
  \adp{32}{1983}{53}.

\bibitem{Bernard}
  D.~Bernard,
   {\it (Perturbed) conformal field theory applied to 2D disordered systems:  An
  introduction},
  \hepth{9509137}.

\bibitem{SchSal}
  V.~Schomerus and H.~Saleur,
  {\it The $GL(1|1)$ WZW model: From supergeometry to logarithmic CFT},
  \npb{734}{2006}{221} [\hepth{0510032}].

\bibitem{Dorn:1994xn}
  H.~Dorn and H.~J.~Otto,
  {\it Two and three point functions in Liouville theory},
  \npb{429}{1994}{375} [\hepth{9403141}].

\bibitem{Zamolodchikov:1995aa}
  A.~B.~Zamolodchikov and A.~B.~Zamolodchikov,
  {\it Structure constants and conformal bootstrap in Liouville field theory},
  \npb{477}{1996}{577} [\hepth{9506136}].

\bibitem{Ponsot:1999uf}
  B.~Ponsot and J.~Teschner,
  {\it Liouville bootstrap via harmonic analysis on a noncompact quantum group},
  \hepth{9911110}.

\bibitem{Teschner:2001rv}
  J.~Teschner,
  {\it Liouville theory revisited},
  \cqg{18}{2001}{R153} [\hepth{0104158}].

\bibitem{Teschner:2003en}
  J.~Teschner,
  {\it A lecture on the Liouville vertex operators},
  \ijmpa{19S2}{2004}{436} [\hepth{0303150}].

\bibitem{Teschner:1997ft}
  J.~Teschner,
  {\it On structure constants and fusion rules in the 
       $SL(2,{\mathbb C})/SU(2)$ WZNW  model},
  \npb{546}{1999}{390} [\hepth{9712256}].

\bibitem{Teschner:1999ug}
  J.~Teschner,
  {\it Operator product expansion and factorization in the $H_3^+$ WZNW model},
  \npb{571}{2000}{555} [\hepth{9906215}].

\bibitem{Teschner:2001gi}
  J.~Teschner,
  {\it Crossing symmetry in the $H_3^+$ WZNW model},
  \plb{521}{2001}{127} [\hepth{0108121}].

\bibitem{Ribault:2005wp}
  S.~Ribault and J.~Teschner,
  {\it $H_3^+$ WZNW correlators from Liouville theory},
  \jhep{06}{2005}{014} [\hepth{0502048}].

\bibitem{Stoyanovsky}
  A.~V.~Stoyanovsky,
  {\it A relation between the Knizhnik--Zamolodchikov and
  Belavin--Polyakov--Zamolodchikov systems of partial differential equations},
 {\tt math-ph/0012013}.

\bibitem{Sklyanin:1987ih}
  E.~K.~Sklyanin,
  {\it Separation of variables in the Gaudin model},
  {\it J.\ Sov.\ Math.} {\bf 47} (1989) 2473
  [{\it Zap.\ Nauchn.\ Semin.} {\bf 164} (1987) 151].

\bibitem{HS}
  Y.~Hikida and V.~Schomerus,
  {\it $H^+_3$ WZNW model from Liouville field theory},
  \jhep{10}{2007}{064} [\arXivid{0706.1030}].

\bibitem{Bouwknegt:1992wg}
  P.~Bouwknegt and K.~Schoutens,
  {\it W symmetry in conformal field theory},
  \prep{223}{1993}{183} [\hepth{9210010}].

\bibitem{QuSc}
  T.~Quella and V.~Schomerus,
  {\it Free fermion resolution of supergroup WZNW models},
  \jhep{09}{2007}{085} [\arXivid{0706.0744}].
  
\bibitem{Bershadsky:1989tc}
  M.~Bershadsky and H.~Ooguri,
  {\it Hidden $OSp(N,2)$ symmetries in superconformal field theory},
  \plb{229}{1989}{374}.  

\bibitem{FH}
  T.~Fukuda and K.~Hosomichi,
  {\it Super Liouville theory with boundary},
  \npb{635}{2002}{215} [\hepth{0202032}].

\bibitem{Saleur:2003zm}
  H.~Saleur and B.~Wehefritz Kaufmann,
  {\it Integrable quantum field theories with supergroup 
  symmetries: The $OSP(1/2)$ case},
  \npb{663}{2003}{443} [\hepth{0302144}].

\bibitem{Essler:2005ag}
  F.~H.~L.~Essler, H.~Frahm and H.~Saleur,
  {\it Continuum limit of the integrable $sl(2/1)$ $3-\bar{3}$ superspin chain},
  \npb{712}{2005}{513} [\condmat{0501197}].

\bibitem{Bernard:2000vc}
  D.~Bernard and A.~LeClair,
  {\it Spin-charge separation and the spin quantum Hall effect},
  \prb{64}{2001}{045306} [\condmat{0003075}].
  
\bibitem{Bhaseen:2000mi}
  M.~J.~Bhaseen, J.~S.~Caux, I.~I.~Kogan and A.~M.~Tsvelik,
  {\it Disordered Dirac fermions: The marriage of three 
  different approaches},
  \npb{618}{2001}{465} [\condmat{0012240}].

\bibitem{LeClair:2007aj}
  A.~LeClair,
  {\it The $gl(1|1)$ super-current algebra: The role of twist and logarithmic
  fields},
  \arXivid{0710.2906}.

\bibitem{df}
  P.~Di~Francesco, P.~Mathieu and D.~S\'{e}n\'{e}chal,
  {\it Conformal field theory},
  Springer (1997).

\bibitem{BPZ}
  A.~A.~Belavin, A.~M.~Polyakov and A.~B.~Zamolodchikov,
  {\it Infinite conformal symmetry in two-dimensional quantum field theory},
  \npb{241}{1984}{333}.

\bibitem{Alekseev:1998mc}
  A.~Y.~Alekseev and V.~Schomerus,
  {\it D-branes in the WZW model},
  \prd{60}{1999}{061901} [\hepth{9812193}].

\bibitem{Creutzig:2007jy}
  T.~Creutzig, T.~Quella and V.~Schomerus,
  {\it Branes in the $GL(1|1)$ WZNW-Model},
  \arXivid{0708.0583}.

\bibitem{Ahn:2002ev}
  C.~Ahn, C.~Rim and M.~Stanishkov,
  {\it Exact one-point function of ${\cal N} = 1$ super-Liouville 
  theory with boundary},
  \npb{636}{2002}{497} [\hepth{0202043}].

\bibitem{HR}
  K.~Hosomichi and S.~Ribault,
  {\it Solution of the $H^+_3$ model on a disc},
  \jhep{01}{2007}{057} [\hepth{0610117}].

\bibitem{Fateev:2007wk}
  V.~Fateev and S.~Ribault,
  {\it Boundary action of the $H_3^+$ model},
  \arXivid{0710.2093}.

\bibitem{Ahn:2002sx}
  C.~Ahn, C.~Kim, C.~Rim and M.~Stanishkov,
  {\it Duality in ${\cal N} = 2$ super-Liouville theory},
  \prd{69}{2004}{106011} [\hepth{0210208}].

\bibitem{Hosomichi:2004ph}
  K.~Hosomichi,
  {\it ${\cal N} = 2$ Liouville theory with boundary},
  \jhep{12}{2006}{061} [\hepth{0408172}].

\bibitem{Saleur:2006tf}
  H.~Saleur and V.~Schomerus,
  {\it On the $SU(2|1)$ WZNW model and its statistical mechanics applications},
  \npb{775}{2007}{312} [\hepth{0611147}].

\bibitem{Maassarani:1996jn}
  Z.~Maassarani and D.~Serban,
  {\it Non-unitary conformal field theory and logarithmic operators for
  disordered systems},
  \npb{489}{1997}{603} [\hepth{9605062}].

\bibitem{Ludwig:2000em}
  A.~W.~W.~Ludwig,
  {\em A free field representation of the $Osp(2|2)$ current 
  algebra at level  $k = -2$, and Dirac fermions in a random 
  $SU(2)$ gauge potential},
  \condmat{0012189}.

\bibitem{Forgacs:1989ac}
  P.~Forgacs, A.~Wipf, J.~Balog, L.~Feher and L.~O'Raifeartaigh,
  {\it Liouville and Toda theories as conformally reduced WZNW theories},
  \plb{227}{1989}{214}.

\bibitem{FSS}
  L.~Frappat, P.~Sorba and A.~Sciarrino,
  {\it Dictionary on Lie superalgebras},
  \hepth{9607161}.

\bibitem{RS}
  R.~C.~Rashkov and M.~Stanishkov,
  {\it Three-point correlation functions in ${\cal N}=1$ 
  super Liouville theory},
  \plb{380}{1996}{49} [\hepth{9602148}].

\bibitem{Poghosian}
  R.~H.~Poghosian,
  {\it Structure constants in the ${\cal N} = 1$ super-Liouville field theory},
  \npb{496}{1997}{451} [\hepth{9607120}].


\end{thebibliography}
\end{document}